\documentclass[11pt,a4paper]{article}
\usepackage[left=2.5cm,top=3cm,bottom=3cm,right=2.5cm]{geometry}
\newcommand{\D}{\mathrm{d}}
\usepackage{amsmath}
\usepackage{amssymb}
\usepackage{cancel}
\usepackage[colorlinks=true
,urlcolor=blue
,anchorcolor=blue
,citecolor=blue
,filecolor=blue
,linkcolor=blue
,menucolor=blue
,linktocpage=true
,pdfproducer=medialab
]{hyperref}
\usepackage{subfigure}
\usepackage{cite}
\usepackage{authblk}
\usepackage[super]{nth}
\usepackage{graphicx}

\renewcommand{\vec}{\boldsymbol}

\newcommand{\M}[2]{\mathcal{M}_{#1}^{(#2)}}
\newcommand{\fM}[2]{\mathcal{M}_{#1}^{(#2)f}}

\newcommand\eik{\mathcal{E}}
\newcommand\ieik{\hat{\mathcal{E}}}
\def\xc{\xi_{c}} 
\newcommand{\cdis}[2][c]{\left(\frac{1}{#2}\right)_{\hspace*{-3pt}#1}}

\newcommand{\bit}[1]{\D\sigma^{(#1)}}
\newcommand{\bbit}[2]{\D\sigma^{(#1)}_{#2}}
\newcommand\MS{\overline{\rm MS}}

\def\pref#1{%
 \ifnum0<0#1\relax
   \newcount\foo%
   \foo=0%
   \loop
     \advance\foo +1
     \D\Upsilon_{\the\foo}
   \ifnum\foo<#1
   \repeat
 \else
   \prod_{i=1}^{#1}\D\Upsilon_i
 \fi
  \D \Phi_{n,#1}
}

\newmuskip\pFqmuskip

\newcommand*\pFq[6][8]{%
  \begingroup % only local assignments
  \pFqmuskip=#1mu\relax
  \mathchardef\normalcomma=\mathcode`,
  \mathcode`\,=\string"8000
  \begingroup\lccode`\~=`\,
  \lowercase{\endgroup\let~}\pFqcomma
  {}_{#2}F_{#3}{\left[\genfrac..{0pt}{}{#4}{#5};#6\right]}%
  \endgroup
}
\newcommand{\pFqcomma}{{\normalcomma}\mskip\pFqmuskip}
\title{A  subtraction scheme for massive QED}
\author[a,b]{T. Engel}
\author[a,b]{A. Signer}
\author[a,b]{Y. Ulrich}
\affil[a]{\sl\small Paul Scherrer Institut\protect\\
CH-5232 Villigen PSI, Switzerland\vspace{0.5cm}}

\affil[b]{\sl\small
Physik-Institut, Universit\"at Z\"urich, \protect\\
Winterthurerstrasse 190,
CH-8057 Z\"urich, Switzerland
}
\begin{document}

\begin{flushright}
PSI-PR-19-19\\
ZU-TH 43/19\\
\today\\
\end{flushright}
{\let\newpage\relax\maketitle}

\vspace{2ex}
\begin{center}
\begin{minipage}[]{0.9\textwidth} {} 
{\sc Abstract:} We present an extension of the FKS subtraction scheme
beyond next-to-leading order to deal with soft singularities in fully
differential calculations within QED with massive fermions. After a
detailed discussion of the next-to-next-to-leading order case, we show
how to extend the scheme to even higher orders in perturbation
theory. As an application we discuss the computation of the
next-to-next-to-leading order QED corrections to the muon decay and
present differential results with full electron mass dependence.

\end{minipage}
\end{center}

\newpage

%%%%%%%%%%%%%%%%%%%%%%%%%%%%%%%%%%%%%%%%%%%%%%%%%%%%%%%%%%%%%%%%%%%%%
\section{Introduction}
%%%%%%%%%%%%%%%%%%%%%%%%%%%%%%%%%%%%%%%%%%%%%%%%%%%%%%%%%%%%%%%%%%%%%

One of the crucial ingredients needed for perturbative calculations in QED
and QCD is a method to perform the usually divergent phase-space
integration. Since we are often interested in matching experimental
procedures as closely as possible, it is essential to be able to
compute fully differential cross sections with several cuts
applied. This renders the phase-space integration too complicated for
analytic evaluation. 

One possibility to perform the divergent phase-space integrations
numerically is through universal infrared subtraction schemes. These
methods allow the calculation of the real corrections at
next-to-leading order (NLO), at least in principle, for any process in
QED or QCD.  This is accomplished by the construction of a counterterm
that will point-wise subtract the singularities in the integrand such
that a numerical integration over the phase space is possible. The
counterterm has to be in a form that allows analytic integration. By
exploiting the universal structure of soft and collinear
singularities, these counterterms can be constructed in a
process-independent way.

Two of the most widely used truly universal schemes at NLO are
FKS~\cite{Frixione1995Three-jet, Frederix2009Automation} and the
dipole formalism~\cite{Catani:1996vz, Catani:2002hc}. FKS treats soft
and collinear singularities separately by constructing different local
counterterms for each. The dipole formalism does away with this
distinction by using Lorentz invariant dipole terms that remove both
singularities simultaneously.

Recent years have seen a huge effort towards the development of universal
schemes for next-to-next-to-leading order (NNLO) calculations (see
e.g.~\cite{ GehrmannDeRidder:2005cm, Catani:2007vq, Czakon:2010td,
  Boughezal:2011jf, Currie:2013vh, Cacciari:2015jma, Caola:2017dug,
  Magnea:2018hab}).  Typically, these schemes were designed with QCD
calculations in mind. This is both their greatest strength and their
greatest weakness: by being applicable to non-Abelian theories, they
all sacrifice simplicity to some extent. This makes them awkward to
use for calculations in massive QED because they effectively treat
collinear singularities that are not present.

Hence, we present the subtraction scheme FKS$^2$ that, while limited
to massive QED, does not suffer from this problem and is very simple
to implement. FKS$^2$ is a natural extension of the FKS scheme to NNLO
for double-soft singularities. It can also readily be extended to even
higher orders in perturbation theory.

Even though the electromagnetic coupling $\alpha$ is much smaller than
the strong coupling, two-loop calculations in QED can be important in
cases where a very high precision is required. As an example we
mention Bhabha scattering (see \cite{Actis:2010gg} and references
therein), which is used for a luminosity measurement at lepton
colliders. Another potential application is muon-electron scattering
which can be used for an alternative determination of the hadronic
contribution to the running of $\alpha$~\cite{Calame:2015fva}.
Recently, also NNLO corrections due to emission from the electron line
for electron-proton scattering have been
computed~\cite{Bucoveanu:2018soy}.

In addition to fixed-order contributions it is often also required to
include multiple soft and/or collinear emissions of photons. Typically
this is done by combining a parton-shower approach with fixed-order
computations~\cite{Barberio:1993qi, Baur:1998kt, CarloniCalame:2003ux,
Balossini:2006wc, Hamilton:2006xz, Schonherr:2008av, Jadach:1995nk}.
This allows to resum logarithmically enhanced terms. A recent example
where resummation is combined with fixed-order NNLO contributions in
QED can be found in \cite{Krauss:2018djz}. While we will not address
resummation directly, it is important to keep it in mind when
constructing a subtraction scheme.

This paper is structured as follows: In Section~\ref{sec:nlo} we
briefly review the FKS scheme at NLO to introduce our notation and
familiarise the reader with the necessary concepts. Next, in
Section~\ref{sec:nnlo} we present FKS$^2$, the extension of the scheme
to NNLO. Further extensions beyond NNLO are discussed in
Section~\ref{sec:beyond}, while referring details of the N$^3$LO case
to Appendix~\ref{app:nnnlo}. In addition, we comment on some generic
properties of the scheme in Section~\ref{sec:comments}.  Next, we use
Section~\ref{sec:results} to demonstrate the validity of FKS$^2$ by
calculating as an example the NNLO QED corrections to the muon decay
in a fully differential way.  Finally, we conclude in
Section~\ref{sec:conclusion}.

%%%%%%%%%%%%%%%%%%%%%%%%%%%%%%%%%%%%%%%%%%%%%%%%%%%%%%%%%%%%%%%%%%%%%
\section{Notation and concepts}\label{sec:nlo}
%%%%%%%%%%%%%%%%%%%%%%%%%%%%%%%%%%%%%%%%%%%%%%%%%%%%%%%%%%%%%%%%%%%%%

Beyond leading order, a physical cross section is computed as a sum of
several separately divergent parts. As a concrete example we consider
a NNLO contribution to a $n$-particle process, which can be written as
\begin{align}
\label{eq:sigmannlo}
    \sigma^{(2)} = \int \Big(\bbit{2}{vv} + \bbit{2}{rv} +
    \bbit{2}{rr} \Big)
=
\int\D\Phi_n\,\M n2
+\int\D\Phi_{n+1}\, \M{n+1}1 +\int\D\Phi_{n+2}\, \M{n+2}0\,.
\end{align}
The double-virtual corrections are obtained by integrating $\M n2$
over the Born phase space $\D\Phi_n$. Here $\M n2$ contains all terms
of the $n$-particle (renormalised) matrix element squared with two
additional powers of the coupling $\alpha$. This includes the
interference term of the two-loop amplitude with the tree-level
amplitude as well as the one-loop amplitude squared. Similarly, the
real-virtual contribution is obtained by integration of $\M{n+1}1$,
the interference of the (renormalised) $(n+1)$-particle one-loop
amplitude with the corresponding tree-level amplitude, over the
$(n+1)$-particle phase space $\D\Phi_{n+1}$. Finally, for the
double-real contribution the tree-level matrix element with two
additional particles, $\M{n+2}0$, is integrated over the corresponding
phase space. In \eqref{eq:sigmannlo} we implicitly assume the presence
of the flux factor (or the analogous factor for a decay rate) as well
as a measurement function that defines the observable in terms of the
particle momenta. The measurement function has to respect infrared
safety, i.e. the observable it defines must not depend on whether or
not one or more additional soft photons are present as arguments of
this function.

In order to have sufficient flexibility in defining the observable,
these phase-space integrals have to be done numerically. However, the
presence of infrared singularities make a direct integration
impossible. In dimensional regularisation with $d=4-2\epsilon$
this would lead to $1/\epsilon$ poles. Instead, a suitable
subtraction has to be made such that the numerical integration is
carried out only with expressions that neither contain implicit soft
singularities from real emissions nor explicit $1/\epsilon$
singularities from loop integrations. Since we are dealing with QED
processes with massive fermions there are no collinear
singularities. Put differently, the collinear poles $1/\epsilon$ are
replaced by $\ln(m)$ terms, where $m$ is a fermion mass. This
considerably simplifies the subtraction procedure and in what follows
we present a scheme that is tailored to this situation.

The structure of soft singularities in QED has been studied a long
time ago by Yennie, Frautschi and Suura (YFS)~\cite{Yennie:1961ad} to
all orders in $\alpha$. The key feature is that after splitting the
amplitude (squared) into a contribution containing the soft
singularity and a contribution free of soft singularities, the former
exponentiate. Thus we can write
\begin{align}
\label{eq:yfs}
\sum_{\ell = 0}^\infty \M{n}{\ell} = 
e^{-\alpha \ieik}\, \sum_{\ell = 0}^\infty \fM{n}{\ell} \,
\end{align}
where $\fM{n}{\ell}$ are free from infrared poles and all
singularities are contained in the eikonal $\ieik$. This does not yet
completely define $\ieik$ as there is some freedom to include finite
terms. The precise definition we will use will be given in
\eqref{eq:inteik}.

The simple structure of \eqref{eq:yfs} can be exploited to resum
leading logarithmic corrections and even combine
this with fixed-order computations. Indeed, there is a long history of
using the YFS approach to construct Monte Carlo
algorithms~\cite{Jadach:1995nk, Hamilton:2006xz, Schonherr:2008av} to
include QED effects in scattering processes.  We will focus on a
fixed-order approach and use the YFS formalism to extend the FKS
subtraction scheme~\cite{Frixione1995Three-jet,
  Frederix2009Automation} to deal with soft singularities in QED
beyond NLO.

%%%%%%%%%%%%%%%%%%%%
\subsection{FKS for soft singularities at NLO}
%%%%%%%%%%%%%%%%%%%%

In this subsection we will briefly summarise the necessary aspects of
the FKS scheme at NLO. Because we only treat soft singularities, FKS
is dramatically simplified. The NLO correction to a cross section is
split into virtual and real parts
\begin{align}
\label{eq:sigmanlo}
    \sigma^{(1)} = \int \Big(\bbit{1}{v} + \bbit{1}{r}  \Big)
= \int\D\Phi_n\,\M n1 +\int\D\Phi_{n+1}\, \M{n+1}0\,.
\end{align}
The real corrections
\begin{align}
\label{eq:nloreal}
\bbit{1}{r} = \D\Phi_{n+1}\, \M{n+1}0\, 
\end{align}
are obtained by integrating the tree-level matrix element $\M{n+1}0$
over the phase space $\D\Phi_{n+1}$. To simplify the discussion we
assume that in the tree-level process described by $\M{n}{0}$ no
final-state photons are present. Hence, in $\M{n+1}0$ only the
particle (photon) with label $n+1$ can potentially become soft. If
there are additional photons (i.e. photons in the tree-level process)
the combinatorics becomes slightly more involved, but the essential
part of the discussion is not affected.

When computing a cross section in the centre-of-mass frame, we choose
coordinates where the beam axis is in $z$ direction. Further, we
denote the (partonic) centre-of-mass energy by $\sqrt s$.  When
computing a decay width we instead parametrise one of the outgoing
particles in $z$ direction and, if necessary, rotate the coordinate
system afterwards.

Following~\cite{Frixione1995Three-jet} we parametrise the momentum of
the additionally radiated particle $n+1$ as\footnote{Note that this
  parametrisation could also tackle initial-state collinear
  singularities because $y$ corresponds to the angle between the
  photon and the incoming particles. However, a different
  parametrisation may be sensible (and is allowed here) to better
  account for pseudo-collinear singularities from light particles (cf.
  Section~\ref{sec:pcs}). What is important in the following is that
  the scaled energy $\xi_1$ is chosen as a variable in the
  parametrisation to ensure a consistent implementation of the
  distributions defined in \eqref{eq:xidist}. }
\begin{align}
\label{eq:kdef}
k_1 = p_{n+1} = \frac{\sqrt s}2\xi_1   (1,\sqrt{1-y_1  ^2}\vec e_\perp,y_1  )\,,
\end{align}
where $\vec e_\perp$ is a $d-2$ dimensional unit vector and the range
of $y_1$ (the cosine of the angle) and $\xi_1$ (the scaled energy) are
$-1\le y_1 \le 1$ and $0\le\xi_1 \le\xi_\text{max}$, respectively. The
upper bound $\xi_\text{max}$ depends on the masses of the outgoing
particles.  Following \cite{Frederix2009Automation} we find
\begin{align*}
\xi_\text{max} = 1-\frac{\Big(\sum_i m_i\Big)^2}{s}\,.
\end{align*}
Further kinematic constraints are assumed to be implemented through
the measurement function.  We write the single-particle phase-space
measure for particle $n+1$ as
\begin{align}
\D\phi_1 &\equiv
   \mu^{4-d}\frac{\D^{d-1}k_1}{(2\pi)^{d-1}\,2k_1^0}
   = \frac{\mu^{2\epsilon}}{2(2\pi)^{d-1}}
   \left(\frac{\sqrt{s}}2\right)^{d-2}
  \xi_1^{1-2\epsilon}(1-y_1^2)^{-\epsilon}\,
      \D\xi_1\,\D y_1\,\D\Omega_1^{(d-2)} =
      \D\Upsilon_1 \D\xi_1 \ \xi_1^{1-2\epsilon}
      \, ,
\label{eq:para}
\end{align}
where the $d-3$ angular integrations $\D\Omega_1^{(d-2)}$ and trivial
factors are collected in $\D\Upsilon_1$.  Denoting by $\D \Phi_{n,1}$
the remainder of the $(n+1)$-parton phase space, i.e. $\D \Phi_{n+1} =
\D \Phi_{n,1} \D\phi_1$, we write the real part of the NLO differential
cross section as
\begin{align}
\bbit{1}{r} &= \D \Phi_{n,1}\,  \D\phi_1\,\M{n+1}0
= \pref1\,  \D\xi_1\ \xi_1^2\M{n+1}0 \xi_1^{-1-2\epsilon} \,.
\label{eq:realnlo}
\end{align}
To isolate the soft singularities in the phase-space integration we
use the identity
\begin{align}\label{eq:xidist}\begin{split}
\xi^{-1-2\epsilon} &=
  -\frac{\xc^{-2\epsilon}}{2\epsilon}\delta(\xi) +
  \cdis{\xi^{1+2\epsilon}}
\,,\\
\left\langle \cdis{\xi^n}, f\right\rangle
&=
\int_0^1\D\xi\,\frac{f(\xi)-f(0)\theta(\xc-\xi)}{\xi^n}
\,,
\end{split}\end{align}
to expand $\xi_1^{-1-2\epsilon}$ in terms of distributions.  Here we
have introduced an unphysical free parameter $\xc$ that can be chosen
arbitrarily~\cite{Frixione1995Three-jet,Frederix2009Automation} as
long as
\begin{align*}
0<\xc\le\xi_\text{max}\,.
\end{align*}
The dependence of $\xc$ has to drop out exactly since no
approximation was made.  Therefore, any fixed value could be
chosen. However, keeping it variable is useful to test the
implementation of the scheme.

Using~\eqref{eq:xidist} we split the real cross section into a hard
and a soft part\footnote{In~\cite{Frixione1995Three-jet} the second
  term is called $\bit{ns}$ for `non-soft'. We will label it $h$ (for
  `hard') instead to avoid confusion when we need more than one such
  label later.}
\begin{subequations}
\begin{align}
\bbit{1}{r} &= \bbit{1}{s}(\xc) + \bbit{1}{h}(\xc)
\,,\\
\bbit{1}{s}(\xc)
 &=
  -\pref1\ \frac{\xc^{-2\epsilon}}{2\epsilon}\ \delta(\xi_1)\,
      \D\xi_1\,
  \Big(\xi_1^2\M{n+1}0\Big)
\,,\\
\label{eq:nloh}
\bbit{1}{h}(\xc)
 &=
 +\pref1\ \cdis{\xi_1^{1+2\epsilon}}
      \D\xi_1
      \Big(\xi_1^2\M{n+1}0\Big)\,.
\end{align}
\end{subequations}
In $\bbit{1}{s}$ we can now (trivially) perform the $\xi_1$
integration.  To do this systematically, we define for photons the
general soft limit $\mathcal{S}_i$ of the $i$-th particle
\begin{align*}
\mathcal{S}_i\M m0\equiv\lim_{\xi_i\to0}\xi_i^2\M m0
    =\eik_i\M{m-1}0
\qquad\text{with}\qquad
\xi_i = \frac{2E_i}{\sqrt{s}}\,,
\end{align*}
where $\M{m-1}0$ is the matrix element for the process without
particle $i$. The eikonal factor 
\begin{align}
\label{eq:eikonal}
    \eik_{i} \equiv \sqrt{4\pi \alpha} \, \sum_{j,k} \frac{p_j\cdot
      p_k}{p_j\cdot n_i\,p_k\cdot n_i}\ {\rm sign}_{jk}
    \qquad\text{with}\qquad p_i=\xi_i n_i\,
\end{align}
is assembled from self- and mixed-eikonals and ${\rm sign}_{jk} =
(-1)^{n_{jk}+1}$, where $n_{jk}$ is the number of incoming particles
or outgoing antiparticles among the particles $i$ and $j$.  Further,
we define the integrated eikonal
\begin{align}
\label{eq:inteik}
\ieik(\xc)
\equiv -\frac{\xc^{-2\epsilon}}{2\epsilon} \int\D\Upsilon_i\ \eik_i
= \xc^{-2\epsilon}\ \ieik(1)
= \sum_{j,k} \ieik_{jk}(\xc)\,,
\end{align}
completing the definition in~\eqref{eq:yfs}.  After $\D \Upsilon_1$
and $\D\xi_1$ integration (under which $\D\Phi_{n,1} \to \D \Phi_n$)
we obtain
\begin{align}
\bbit{1}{s}(\xc)\
    & \stackrel{\int\D\Upsilon_1\D\xi_1}{\longrightarrow}
\
\D \Phi_{n}\ \ieik(\xc)\,\M n0\,.
\label{eq:nlo:s}
\end{align}
This part now contains explicit $1/\epsilon$ poles that cancel against
poles in the virtual cross section. The second term of the real
corrections, $\bbit{1}{h}$ given in \eqref{eq:nloh}, is finite and can
be integrated numerically after setting $d=4$.  Combining the real and
virtual corrections, the NLO correction is given by
\begin{subequations}
\label{eq:nlo:4d}
\begin{align}
%\begin{split}
\sigma^{(1)} &=
\sigma^{(1)}_n(\xc) + \sigma^{(1)}_{n+1}(\xc) \, , \\
\sigma^{(1)}_n(\xc) &= \int
\ \D\Phi_n^{d=4}\,\Bigg(
    \M n1
   +\ieik(\xc)\,\M n0
\Bigg) = 
\int \ \D\Phi_n^{d=4}\, \fM n1
\,,
%\end{split}
\label{eq:nlo:n}
\\
\sigma^{(1)}_{n+1}(\xc) &= \int 
\ \D\Phi^{d=4}_{n+1}
  \cdis{\xi_1} \big(\xi_1\, \fM{n+1}0 \big)
\label{eq:nlo:n1}
\, .
\end{align}
\end{subequations}
We have used $\M{n+1}0 = \fM{n+1}0$ and absorbed one of the
$\xi_1$ factors multiplying $\M{n+1}0$ in~\eqref{eq:nloh} in the phase
space $\D\Phi^{d=4}_{n+1}$. Contrary to \eqref{eq:sigmanlo}, there are
no soft singularities present in \eqref{eq:nlo:4d}. According to
\eqref{eq:yfs} the explicit $1/\epsilon$ poles cancel between the two
terms in the integrand of \eqref{eq:nlo:n} and the phase-space
integration in \eqref{eq:nlo:n1} is also manifestly finite.

We note that $\mathcal{S}_i$ is invariant under rotations, but not
Lorentz invariant, because it contains the explicit energy
$E_i$. Hence, also $\eik_i$ and $\ieik$ are only invariant under
rotations but not under general Lorentz transformations. The
integrated eikonal $\ieik_{jk}$ has been computed
in~\cite{Frederix2009Automation}, dropping terms of
$\mathcal{O}(\epsilon)$. As we will see this is sufficient even beyond
NLO. The expression is given in Section~\ref{app:eikonal}, using our
conventions.

%%%%%%%%%%%%%%%%%%%%%%%%%%%%%%%%%%%%%%%%%%%%%%%%%%%%%%%%%%%%%%%%%%%%%
\section{\texorpdfstring{FKS$^2$}{FKS2}: NNLO extension}
\label{sec:nnlo}
%%%%%%%%%%%%%%%%%%%%%%%%%%%%%%%%%%%%%%%%%%%%%%%%%%%%%%%%%%%%%%%%%%%%%

In the following, we discuss the extension of FKS to NNLO, while still
limiting ourselves to massive QED. To simplify the discussion in this
section, we assume that all (suitably renormalised) matrix elements are
known to sufficient order in the coupling and expansion in
$\epsilon$. In Section~\ref{sec:scheme} we will state what precisely
is needed for a NNLO computation.

%%%%%%%%%%%%%%%%%%%%
\subsection{Real-virtual correction}
%%%%%%%%%%%%%%%%%%%%

The treatment of the real-virtual contribution
\begin{align}
\label{eq:nnlorv}
\bbit{2}{rv} = \D\Phi_{n+1}\, \M{n+1}1
\end{align}
proceeds along the lines of normal FKS because it is a
$(n+1)$-particle contribution. Again we assume that there is only one
external particle, with label $n+1$, that can potentially become
soft. We use \eqref{eq:xidist} with another unphysical cut-parameter
$\xi_{c_A}$ to split the real-virtual cross section into a soft and a
hard part
\begin{align}
\label{eq:shnnlo}
\bbit{2}{rv} = \bbit{2}{s}(\xi_{c_A}) + \bbit{2}{h}(\xi_{c_A}) \, .
\end{align}
For $\bbit{2}{s}$ the analogy to the NLO case is particularly strong
because there is no genuine one-loop eikonal
contribution~\cite{Bierenbaum:2011gg,Catani:2000pi}, i.e. the soft
limit of the real-virtual matrix element is
\begin{align*}
\mathcal{S}_{n+1}\M{n+1}1=\eik_{n+1} \M n1\,,
\end{align*}
with the same $\eik_{n+1}$ as in~\eqref{eq:eikonal}. Therefore,
compared to \eqref{eq:nlo:s} the definition of the soft part remains
essentially unchanged
\begin{align}
\bbit{2}{s}(\xi_{c_A})\
    & \stackrel{\int\D\Upsilon_1\D\xi_1}{\longrightarrow}
  \ \D \Phi_{n}\ \ieik(\xi_{c_A})\,\M n1\,.
\label{eq:nnlo:s}
\end{align}
However, $\bbit{2}{s}$ has a double-soft $1/\epsilon^2$ pole from the
overlap of the soft $1/\epsilon$ poles of $\ieik$ and $\M n1$.

Unfortunately, $\bbit{2}{h}$ is not yet finite as it contains an
explicit $1/\epsilon$ pole from the loop integration. To deal with
this pole we use eikonal subtraction, i.e. we split the real-virtual
matrix element according to
\begin{align}
 \label{eq:polemrv}
  \M{n+1}1 \equiv \fM{n+1}1(\xi_{c_B}) - \ieik(\xi_{c_B})\,\M{n+1}0\,
\end{align}
into a finite and a divergent piece.  The pole of $ \M{n+1}1$ is now
contained in the integrated eikonal of $\ieik(\xi_{c_B})\,\M{n+1}0$,
whereas the eikonal-subtracted matrix element $\fM{n+1}1$ is free from
poles. This is again the YFS split, mentioned in \eqref{eq:yfs}. In
\eqref{eq:polemrv} we have introduced yet another initially
independent cut-parameter $\xi_{c_B}$.

\begin{subequations}
With the help of \eqref{eq:polemrv} we can now write
\begin{align}
\label{eq:nnloh}
\begin{split}
\bbit{2}{h}(\xi_{c_A}) &=
  \pref1 \D\xi_1\,
  \cdis[c_A]{\xi_1^{1+2\epsilon}} \big(\xi_1^2 \M{n+1}1\big)
\\&=
\bbit{2}{f}(\xi_{c_A},\xi_{c_B}) + \bbit{2}{d}(\xi_{c_A},\xi_{c_B}) \, ,
\end{split}
\end{align}
where $c_A$ indicates that the subtraction should be performed with
the cut parameter $\xi_{c_A}$. The finite piece
\begin{align}
\bbit{2}{f}(\xi_{c_A},\xi_{c_B}) &=
  \pref1 \D\xi_1\,
  \cdis[c_A]{\xi_1^{1+2\epsilon}} \big(\xi_1^2\fM{n+1}1(\xi_{c_B})\big)
\,,\label{eq:nnlo:fin}
\end{align}
can be integrated numerically with $\epsilon=0$. Integrating the
divergent piece, $\bbit{2}{d}$, over the complete phase space we
obtain
\begin{align}
\begin{split}
\int\bbit{2}{d}(\xi_{c_A},\xi_{c_B}) &=
  -\int\pref1 \D\xi_1\, 
  \cdis[c_A]{\xi_1^{1+2\epsilon}}   \big(
      \ieik(\xi_{c_B})\, \xi_1^2\,\M{n+1}0
  \big)
%\\&
\equiv -\mathcal{I}(\xi_{c_A},\xi_{c_B})
\,,
\end{split}\label{eq:nnlo:sin}
\end{align}
where in $\mathcal{I}$ the first argument refers to the cut-parameter
of the $\xi$ integration and the second to the argument of
$\hat{\mathcal{E}}$. This process- and observable-dependent function
is not finite and generally very tedious to compute. Even for the
simplest cases it gives rise to complicated analytic expressions
including for example Appell's $F_i$ functions. However, as we will
see it is possible to cancel its contribution exactly with the
double-real emission.
\end{subequations}

To summarise, the real-virtual corrections are given by
\begin{align}
\label{eq:rv}
\bbit{2}{rv} &= \bbit{2}{s}(\xi_{c_A})
+ \bbit{2}{f}(\xi_{c_A},\xi_{c_B})
+ \bbit{2}{d}(\xi_{c_A},\xi_{c_B})
\, ,
\end{align}
where the expressions for $\bbit{2}{s}$, $\bbit{2}{f}$ and
$\bbit{2}{d}$ can be read off from \eqref{eq:nnlo:s},
\eqref{eq:nnlo:fin} and \eqref{eq:nnlo:sin}, respectively. We point
out that $\bbit{2}{rv}$ is independent of both $\xi_{c_A}$ and
$\xi_{c_B}$.

%%%%%%%%%%%%%%%%%%%%
\subsection{Double-real correction}
%%%%%%%%%%%%%%%%%%%%

For the double-real contribution
\begin{align}
\label{eq:nnlorr}
\bbit{2}{rr} = \D\Phi_{n+2}\, \M{n+2}0
\end{align} 
we have to consider $\M{n+2}{0}$, the matrix element for the process
with two additional photons (with labels $n+1$ and $n+2$) w.r.t. the
tree-level process. We extend the parametrisation~\eqref{eq:para}
accordingly to
\begin{align}
k_1 = p_{n+1} = \frac{\sqrt s}2\xi_1 (1,\sqrt{1-y_1^2}\vec e_\perp,y_1)\,,
&\qquad
k_2 = p_{n+2} = \frac{\sqrt s}2\xi_2 R_\phi(1,\sqrt{1-y_2^2}\vec e_\perp,y_2)\,,
\end{align}
with $-1\le y_i\le 1$, $0\le\xi_i\le\xi_\text{max}$ and a
$(d-2)$-dimensional rotation matrix $R_\phi$. Writing the phase space
as $\D \Phi_{n+2} = \D \Phi_{n,2} \D\phi_1 \D\phi_2$, the double-real
contribution becomes
\begin{align}
\begin{split}
\bbit{2}{rr} &= \D \Phi_{n,2} \D\phi_1 \D\phi_2\,\ \frac1{2!} \M{n+2}0
\\&=
\pref2\ \D\xi_1\,\D\xi_2\, \frac1{2!} \big(\xi_1^2\xi_2^2\M{n+2}0\big)\ 
  \xi_1^{-1-2\epsilon}\,\xi_2^{-1-2\epsilon}\,,
\end{split}\label{eq:nnlo:sym}
\end{align}
where we have used analogous definitions as in \eqref{eq:para} and
\eqref{eq:realnlo}. The only difference between $\D \Phi_{n,1}$ and $\D
\Phi_{n,2}$ is in the argument of the $\delta$~function that ensures
momentum conservation.  Note that the factor $1/2!$ is due to the
symmetry of identical particles.

Again, we use \eqref{eq:xidist} with two new cut parameters
$\xi_{c_1}$ and $\xi_{c_2}$ to expand $\bbit{2}{rr}$ in terms of
distributions as
\begin{align}
\label{eq:rr}
&\qquad  \bbit{2}{rr} = \bbit{2}{ss}(\xi_{c_1},\xi_{c_2}) +
\bbit{2}{sh}(\xi_{c_1},\xi_{c_2}) + \bbit{2}{hs}(\xi_{c_1},\xi_{c_2})
+ \bbit{2}{hh}(\xi_{c_1},\xi_{c_2})
\,,\\[10pt] \nonumber
& \left\{\def\arraystretch{1.6}\begin{array}{c}
   \bbit{2}{ss}(\xi_{c_1},\xi_{c_2}) \\[5pt] 
   \bbit{2}{hs}(\xi_{c_1},\xi_{c_2}) \\[5pt]
   \bbit{2}{sh}(\xi_{c_1},\xi_{c_2}) \\[5pt] 
   \bbit{2}{hh}(\xi_{c_1},\xi_{c_2})
\end{array}\right\} =
    \pref2\ \frac1{2!}\ 
\left\{\def\arraystretch{1.9}\begin{array}{c}
   \frac{\xi_{c_1}^{-2\epsilon}}{2\epsilon}\delta(\xi_1) \,
   \frac{\xi_{c_2}^{-2\epsilon}}{2\epsilon}\delta(\xi_2)
   \\
  -\frac{\xi_{c_2}^{-2\epsilon}}{2\epsilon}\delta(\xi_2) \,
   \cdis[c_1]{\xi_1^{1+2\epsilon}}
   \\
  -\frac{\xi_{c_1}^{-2\epsilon}}{2\epsilon}\delta(\xi_1) \,
   \cdis[c_2]{\xi_2^{1+2\epsilon}}
   \\
   \cdis[c_1]{\xi_1^{1+2\epsilon}}\,
   \cdis[c_2]{\xi_2^{1+2\epsilon}}
\end{array}\right\}\, \D\xi_1\,\D\xi_2
\ \xi_1^2\xi_2^2\M{n+2}0\,.
\end{align}
We note that for $\xi_{c_1} = \xi_{c_2} \to \xi_c$ we have 
$\int\bbit{2}{sh}(\xi_{c},\xi_{c}) =
\int\bbit{2}{hs}(\xi_{c},\xi_{c})$.

The contribution from $\bbit{2}{hh}$ can be integrated numerically
with $\epsilon=0$ because it is finite everywhere.

For the mixed contributions  $\bbit{2}{hs}$ and $\bbit{2}{sh}$ we use
\begin{align*}
\mathcal{S}_{i}\M{n+2}0 = \eik_i \M{n+1}0
\qquad\text{with}\qquad
i\in\{n+1,n+2\}\, .
\end{align*}
Considering first $\bbit{2}{hs}$, we perform the $\xi_2$ integration
(under which $\D \Phi_{n,2} \to \D \Phi_{n,1}$) and use \eqref{eq:inteik} to
do the $\D\Upsilon_2$ integration to obtain
\begin{subequations}
\begin{align}
\int\bbit{2}{hs}(\xi_{c_1},\xi_{c_2}) &=
    \int \D\Upsilon_1 \D \Phi_{n,1} \ \frac1{2!}\ \int\D\xi_1\ \cdis[c_1]{\xi_1^{1+2\epsilon}}
    \big(\xi_1^2\M{n+1}0\big)
    \ieik(\xi_{c_2})
%\notag\\&
 = \frac1{2!}\mathcal{I}(\xi_{c_1},\xi_{c_2})
\end{align}
Similarly, we get
\begin{align}
\int\bbit{2}{sh}(\xi_{c_1},\xi_{c_2}) &=
  \frac1{2!}\mathcal{I}(\xi_{c_2},\xi_{c_1})
\,.
\end{align}
\end{subequations}
Thus, we find again the integral $\mathcal{I}$ of \eqref{eq:nnlo:sin}.

Finally, we turn to the double-soft contribution $\bbit{2}{ss}$. Since
\begin{align*}
\Big(\mathcal{S}_i\circ\mathcal{S}_j\Big)\M{n+2}0 
&= \Big(\mathcal{S}_j\circ\mathcal{S}_i\Big)\M{n+2}0
 = \eik_i\eik_j\, \M{n}0
\qquad\text{with}\qquad
i\neq j\in\{n+1,n+2\}\,,
\end{align*}
the $\xi$ integrals in $\bbit{2}{ss}$ factorise. Therefore, we can do
the $\xi_1, \Upsilon_1$ integrations independently from the $\xi_2,
\Upsilon_2$ integrations and obtain
\begin{align}
\begin{split}
\bbit{2}{ss}(\xi_{c_1},\xi_{c_2}) \ 
&\stackrel{\int\D\Upsilon_{1,2}\D\xi_{1,2}}{\longrightarrow} \
\D \Phi_{n}\ \frac1{2!}\
\ieik(\xi_{c_1})\ieik(\xi_{c_2})\,\M{n}0\, .
\end{split}\label{eq:nnlo:ss}
\end{align}
It is clear that the simplicity of the infrared structure of QED with
massive fermions is crucial for reducing the complexity of the
procedure described in the steps above.

\clearpage
%%%%%%%%%%%%%%%%%%%%
\subsection{Combination}
%%%%%%%%%%%%%%%%%%%%

At this stage we have introduced four different cutting parameters
$\xi_{c_A}$ and $\xi_{c_B}$ as well as $\xi_{c_1}$ and $\xi_{c_2}$.
All of these are unphysical, arbitrary parameters that can take any
value $0<\xi_{c_i}\le\xi_\text{max}$.  In total we have to deal with
seven different contributions. Two of them, $\bbit{2}{s}$ and
$\bbit{2}{ss}$, are very simple as they just depend on the
eikonal. Another two contributions $\bbit{2}{f}$ and $\bbit{2}{hh}$
can be calculated numerically with $\epsilon=0$.

The sum of the three remaining auxiliary contributions $\bbit{2}{d}$,
as well as $\bbit{2}{sh}$ and $\bbit{2}{hs}$, only depend on the
function $\mathcal{I}$ defined above
\begin{align}
\nonumber
\int \bbit{2}{aux}(\{\xi_{c_i}\}) &\equiv
\int \Big( \bbit{2}{d}(\xi_{c_A},\xi_{c_B})
+\bbit{2}{hs}(\xi_{c_1},\xi_{c_2})+\bbit{2}{sh}(\xi_{c_1},\xi_{c_2}) \Big)
\\
\label{eq:aux}
&=
-\mathcal{I}(\xi_{c_A},\xi_{c_B})
+\frac1{2!}\mathcal{I}(\xi_{c_1},\xi_{c_2})
+\frac1{2!}\mathcal{I}(\xi_{c_2},\xi_{c_1})
\,.
\end{align}
Note that, due to the sign difference and the symmetry factor,
$\bbit{2}{aux}$ vanishes if we choose 
\begin{align}
\label{eq:allxi}
\xc \equiv \xi_{c_A} = \xi_{c_B} = \xi_{c_1} = \xi_{c_2}\,.
\end{align}
This cancellation will not be affected by the measurement
function. Thus, in what follows we will make the choice
\eqref{eq:allxi}, avoiding the computation of the potentially
difficult $\mathcal{I}$ function.\footnote{It is possible to compute
  the auxiliary contribution $\bbit{2}{aux}$ numerically keeping all
  $\xi_{c_i}$ different. While this complicates the implementation of
  the scheme it can be helpful to validate the code, see
  Section~\ref{sec:xic}.} 

We can now collect the non-vanishing contributions, sorted by
remaining integrations
\begin{subequations}
\begin{align}
\sigma^{(2)} &= \sigma^{(2)}_n    (\xc) + 
          \sigma^{(2)}_{n+1} (\xc) + 
          \sigma^{(2)}_{n+2} (\xc)
\,,\\
\label{eq:nnloind0}
\sigma^{(2)}_n (\xc) &=
\int\Big(
\D\Phi_n\, \M n2  +
    \bbit{2}{s}+\bbit{2}{ss} \Big)
\,,\\
\label{eq:nnloind1}
\sigma^{(2)}_{n+1} (\xc) &= \int \bbit{2}{f} =\int
 \pref1 \D\xi\, 
  \cdis{\xi^{1+2\epsilon}} \, \xi^2\fM{n+1}1(\xc)
\,,\\
\label{eq:nnloind2}
\sigma^{(2)}_{n+2} (\xc) &= \int \bbit{2}{hh}
     = \int \pref2 \D\xi_1 \D\xi_2  \frac1{2!}
   \cdis{\xi_1^{1+2\epsilon}}\,
   \cdis{\xi_2^{1+2\epsilon}}\,
     \Big(\xi_1^2\xi_2^2\M{n+2}0\Big)\, .
\end{align}
\end{subequations}
The three terms of the integrand of $\sigma^{(2)}_n$ are separately
divergent. However, in the sum the $1/\epsilon$ poles cancel. The
other parts, $\sigma^{(2)}_{n+1}$ and $\sigma^{(2)}_{n+2}$, are finite
by construction. Hence, we can set $d=4$ everywhere (except in the
individual pieces of the integrand of $\sigma^{(2)}_n$) and obtain
\begin{subequations}
\label{eq:nnlo:4d}
\begin{align}
\begin{split}
\sigma^{(2)}_n(\xc) &= \int
\ \D\Phi_n^{d=4}\,\bigg(
    \M n2
   +\ieik(\xc)\,\M n1
   +\frac1{2!}\M n0 \ieik(\xc)^2
\bigg) = 
\int \ \D\Phi_n^{d=4}\, \fM n2
\,,
\end{split}\label{eq:nnlo:n}
\\
\sigma^{(2)}_{n+1}(\xc) &= \int 
\ \D\Phi^{d=4}_{n+1}
  \cdis\xi \Big(\xi\, \fM{n+1}1(\xc)\Big)
\,,\\
\sigma^{(2)}_{n+2}(\xc) &= \int
\ \D\Phi_{n+2}^{d=4}
   \cdis{\xi_1}\,
   \cdis{\xi_2}\,
     \Big(\xi_1\xi_2\, \fM{n+2}0\Big) \, .
\end{align}
\end{subequations}
This is the generalisation of \eqref{eq:nlo:4d} to NNLO. In the
integrand of \eqref{eq:nnlo:n} the build up of the exponentiated
singular part $e^{\alpha \ieik}$ is recognisable. For $\fM{n}\ell$ to
be finite, $\ieik$ has to contain the soft $1/\epsilon$ pole. However,
any choice of the finite part is possible in principle. We have chosen
to define the finite matrix elements through eikonal subtraction,
\eqref{eq:polemrv}. This ensures that the auxiliary contributions
cancel and the remaining parts $\sigma^{(2)}_{n+1}$ and
$\sigma^{(2)}_{n+2}$ have a very simple form. Terms of
$\mathcal{O}(\epsilon)$ in $\ieik$ have no effect since they do not
modify $\fM{n}\ell$ after setting $d=4$. This means we can set them to
zero and there is no need to compute the integral \eqref{eq:inteik}
beyond finite terms.

%%%%%%%%%%%%%%%%%%%%%%%%%%%%%%%%%%%%%%%%%%%%%%%%%%%%%%%%%%%%%%%%%%%%%
\section{Beyond NNLO} \label{sec:beyond}
%%%%%%%%%%%%%%%%%%%%%%%%%%%%%%%%%%%%%%%%%%%%%%%%%%%%%%%%%%%%%%%%%%%%%

%%%%%%%%%%%%%%%%%%%%
\subsection{\texorpdfstring{FKS$^3$: extension to N$^3$LO}{FKS3: extension to N3LO}}
%%%%%%%%%%%%%%%%%%%%

First steps towards extending universal schemes beyond NNLO have been
made in QCD~\cite{Currie:2018fgr}.  The simplicity of FKS$^2$ suggests
that this paradigm is a promising starting point for further extension
to ${\rm N}^3{\rm LO}$ in massive QED, provided that all matrix
elements are known.

At ${\rm N}^3{\rm LO}$, we have four terms
\begin{align*}
\sigma^{(3)} = \int\D\Phi_{n  } \M{n  }3
       + \int\D\Phi_{n+1} \M{n+1}2
       + \int\D\Phi_{n+2} \M{n+2}1
       + \int\D\Phi_{n+3} \M{n+3}0\,,
\end{align*}
which are separately divergent. In order to reorganise these four
terms into individually finite terms, we repeatedly use
\eqref{eq:xidist} to split the phase-space integrations into hard and
soft and \eqref{eq:polemrv} to split the matrix element into finite
and divergent parts.  In principle we could choose many different
$\xc$ parameters. However, from the experience of FKS$^2$ we expect
decisive simplifications if we choose them all to be the same. Indeed,
as is detailed in Appendix~\ref{app:nnnlo}, there are now three
different auxiliary integrals that enter in intermediate
steps. However, if all $\xc$ parameters are chosen to be equal, their
contributions cancel for any cross section, similar to \eqref{eq:aux}.
Hence, writing 
\begin{align}
\label{eq:nnnlocomb}
\D\sigma^{(3)} &= \bbit{3}{n}(\xc) + \bbit{3}{n+1}(\xc) 
+ \bbit{3}{n+2}(\xc) + \bbit{3}{n+3}(\xc) 
\end{align}
all terms are separately finite and, as discussed in detail in
Appendix~\ref{app:nnnlo}, given by
\begin{subequations}
\label{eq:nnnloparts}
\begin{align}
\bbit{3}{n}(\xc) &=
  \D\Phi_n^{d=4}   \fM{n}3 
\,,\\
\bbit{3}{n+1}(\xc) &= \D\Phi_{n+1}\, \cdis{\xi_1}
                  \Big(\xi_1\, \fM{n+1}2(\xc)\Big)
\,,\\
\bbit{3}{n+2}(\xc) &=  \frac1{2!}\, \D\Phi_{n+2}\, 
   \cdis{\xi_1}\, \cdis{\xi_2}\,
   \Big( \xi_1 \xi_2 \, \fM{n+2}1(\xc) \Big)
\,,\\
\bbit{3}{n+3}(\xc) &= \frac1{3!}\,
  \D\Phi_{n+3}\, 
   \cdis{\xi_1}\,
   \cdis{\xi_2}\,
   \cdis{\xi_3} \ \Big(
   \xi_1\xi_2\xi_3\, \fM{n+3}0(\xc)\Big) \,.
\end{align}
\end{subequations}
Once more we have used the fact that for tree-level amplitudes
$\M{n+3}0 = \fM{n+3}0$.  As always, the $\xc$ dependence cancels between
the various parts such that $\bit{3}$ is independent of this
unphysical parameter.

%%%%%%%%%%%%%%%%%%%%
\subsection{\texorpdfstring{FKS$^\ell$: extension to N$^\ell$LO}{FKSl: extension to NlLO}}
\label{sec:nllo}
%\subsection{${\rm N}^\ell{\rm LO}$ generalisation}
%%%%%%%%%%%%%%%%%%%%

The pattern that has emerged in the previous cases leads to the 
following extension to an arbitrary order $\ell$ in perturbation
theory:
\begin{subequations}
\label{eq:nellocomb}
\begin{align}
\D\sigma^{(\ell)} &= \sum_{j=0}^\ell \bbit{\ell}{n+j}(\xc)\, ,
\\
\bbit{\ell}{n+j}(\xc) &=  \D\Phi_{n+j}^{d=4}\,\frac{1}{j!} \, 
\bigg( \prod_{i=1}^j \cdis{\xi_i} \xi_i \bigg)\,
 \fM{n+j}{\ell-j}(\xc)\,.
\end{align}
\end{subequations}
The eikonal subtracted matrix elements
\begin{align*}
\fM{m}\ell &= \sum_{j=0}^\ell\frac{\ieik^j}{j!} \M{m}{\ell-j}\,,
\end{align*}
(with the special case $\fM{m}0 = \M{m}0$ included) are free from
$1/\epsilon$ poles, as indicated in \eqref{eq:yfs}. Furthermore, the
phase-space integrations are manifestly finite.

%%%%%%%%%%%%%%%%%%%%
\section{Comments on and properties of \texorpdfstring{FKS$^\ell$}{FKSl}}
\label{sec:comments}
%%%%%%%%%%%%%%%%%%%%

With the scheme now established, let us discuss a few non-trivial
properties that are helpful during implementation and testing.

%%%%%%%%%%%%%%%%%%%%
\subsection{Regularisation-scheme and scale dependence}
\label{sec:scheme}
%%%%%%%%%%%%%%%%%%%%

In QED calculations it is advantageous to calculate the matrix
elements $\M{n}{\ell}$ in the on-shell scheme for $\alpha$ (and the
masses). This way the only $\mu$ dependence is in a global prefactor
$\mu^{2\epsilon}$ induced through the integral measure. The same holds
for the integrated eikonal. Hence, for the finite matrix elements
$\fM{n}{\ell}$ there is no $\mu$ dependence after setting $d=4$.  

A similar argument can be made for the regularisation-scheme
dependence. So far we have implicitly assumed that the computation is
performed in conventional dimensional regularisation. However, it is
often more convenient to use other dimensional schemes, where
e.g. external particles are treated in four
dimensions~\cite{Gnendiger:2017pys}. As discussed in
\cite{Broggio:2015dga, Gnendiger:2016cpg}, after proper
renormalisation and infrared subtraction the matrix elements are
regularisation-scheme independent for $d=4$. As there are no collinear
singularities, the eikonal-subtracted $\fM{n}\ell$ are scheme
independent. Moreover, the integrated eikonal is scheme independent.

To be concrete, we list the input that is required for a computation
of a physical cross section at NNLO in QED. The important point is
that once the final expressions for a NNLO cross section,
\eqref{eq:nnlo:4d}, or beyond, \eqref{eq:nellocomb}, are obtained, we
can set $d=4$ everywhere.

\begin{itemize}

    \item
    The two-loop matrix element $\M n2$ is known with non-vanishing
    masses up to $\mathcal{O}(\epsilon^0)$. In general this is a
    bottleneck because the necessary master integrals are only known
    for a very select class of processes, not to mention the algebraic
    complexity. However, it is possible to approximate $\bit{2}$ using
    `massification' of $\M n2$~\cite{Engel:2018fsb, Mitov:2006xs,
      Becher:2007cu} (see Section~\ref{sec:massification}).

    \item
    The renormalised one-loop matrix element $\M n1$ of the
    $n$-particle process is known including $\mathcal{O}(\epsilon^1)$
    terms. This is usually the case for NNLO calculation as it is
    needed for the sub-renormalisation $\M n2 \supset \delta Z\times
    \M n1$ as well as the one-loop amplitude squared, which is part of
    $\M n2$. Once these pieces are assembled to $\fM{n}2$,
    the $\mathcal{O}(\epsilon)$ terms can be dropped.  

  \item
    The renormalised real-virtual matrix element $\M{n+1}1$ is known
    with non-vanishing masses. Terms $\mathcal{O}(\epsilon)$ are not
    required.

  \item
    $\M{n+2}0$ is known in four dimensions. In intermediate steps, the
    matrix elements $\M{n}0$ and $\M{n+1}0$ are required to
    $\mathcal{O}(\epsilon^2)$ and $\mathcal{O}(\epsilon)$,
    respectively. However, depending on the regularisation scheme,
    such terms might actually be absent. In any scheme, once $\fM{n}2$
    and $\fM{n+1}1$ is assembled, the $\mathcal{O}(\epsilon)$ terms
    can be dropped.

\end{itemize}

%%%%%%%%%%%%%%%%%%%%
\subsection{Massification} \label{sec:massification}
%%%%%%%%%%%%%%%%%%%%

The bottleneck in the computation of cross sections for massive QED at
N$^\ell$LO is the availability of the matrix element $\M{n}\ell$. A
potentially interesting option is to study the direct evaluation of
the finite $\fM{n}{\ell}$, opening up the possibility of using
numerical methods. However, the $\M{n}{\ell}$ are traditionally
computed analytically and then combined to $\fM{n}{\ell}$. These
computations are usually much simpler if some (or all) fermion masses,
$m$, are set to zero.  Unfortunately, this also spoils FKS$^\ell$.
However, if $m$ is small compared to the other kinematic quantities,
an option is to start from the massless case and subsequently
`massify' $\fM{n}{\ell}$~\cite{Penin:2005eh, Mitov:2006xs,
Becher:2007cu, Engel:2018fsb}. This converts the collinear
$1/\epsilon$ singularities of $\fM{n}\ell(m=0)$ into $\log(m)$ terms
that will cancel against corresponding `singularities' of the real
corrections.  In addition, it retains in $\bbit{\ell}{n}$ the finite
$\log(m)$ terms that are present in differential distributions.
However, terms $m \log(m)$ that vanish in the limit $m\to 0$ will be
neglected.

It should be noted that a similar procedure in $\bbit{\ell}{n+j}$ is
less straightforward. It is not possible to naively use massification
of $\fM{n+j}{\ell-j}$. The remaining phase-space integration over the
$j$ additional particles requires a non-vanishing $m$ to avoid a
collinear singularity. Thus, using full $m$ dependence in
$\bbit{\ell}{n+j}$, but only partial $m$ dependence in
$\bbit{\ell}{n}$ through a massified $\fM{n}{\ell}$ results in a
mismatch in terms $m \log(m)$. Since the whole procedure of
massification is anyway only correct up to such terms, the
mismatch should not cause additional problems.

%%%%%%%%%%%%%%%%%%%%
\subsection{Phase-space parametrisation}\label{sec:pcs}
%%%%%%%%%%%%%%%%%%%%

A further issue in connection with small lepton masses is related to
the phase-space parametrisation.  The phase space has to be
constructed in any way that allows the distributions to be well
implemented, i.e. $\xi_i$ should be an integration variable of the
numerical integrator. In addition, for small $m$ there are potentially
numerical problems due to pseudo-collinear singularities. In fact,
these regions produce precisely the $\log(m)$ terms that correspond to
the collinear `singularities' of the real part. These $\log(m)$ terms
will cancel the virtual collinear `singularities' mentioned
above. Hence, for small $m$ there is a numerically delicate
cancellation. For the simple observables presented in
Section~\ref{sec:results} this problem could be solved through a
dedicated tuning of the phase-space parametrisation. For more
complicated processes another solution might need to
be implemented in the future. The idea is to subtract such regions
from the integrand and add them back in integrated
form~\cite{Dittmaier:1999mb}. This amounts to actually extend the
subtraction scheme to collinear singularities. It also offers the
possibility to use massification for $\bbit{\ell}{n+j}$. To this end,
$\fM{n+j}{\ell-j}(m=0)$ is massified. The resulting collinear
singularities due to phase-space integration with massless external
fermions are treated as follows: first they are subtracted at
integrand level; second their integrated contribution is added back in
massified form.

%%%%%%%%%%%%%%%%%%%%%%%%%%%%%%%%%%%%%%%%%%%%%%%%%%%%%%%%%%%%%%%%%%%%%
\section{Muon decay} \label{sec:results}
%%%%%%%%%%%%%%%%%%%%%%%%%%%%%%%%%%%%%%%%%%%%%%%%%%%%%%%%%%%%%%%%%%%%%

As an example for the subtraction scheme presented in the
previous section we will discuss the muon decay 
\begin{align}
\label{eq:process}
\mu(p_1) \to e(p_2) \nu\bar\nu + \{\gamma(p_5), \gamma(p_6)\} 
\end{align}
at NNLO in the Fermi theory of weak interactions.\footnote{To compare
  with literature we also have to consider the case of an $\{e^+(p_5),
  e^-(p_6)\}$ pair in the final state.} NLO corrections to the muon
decay have been known for many
decades~\cite{Kinoshita:1958ru,PhysRev.101.866}. Using the optical
theorem, the NNLO QED corrections to the decay width were calculated
around the turn of the millennium assuming vanishing electron
masses~\cite{vanRitbergen:1999fi}. Over the course of the next decade,
the electron energy spectrum, which is not infrared finite in the
limit $m_e\to0$, was calculated. At first, only its logarithms were
known analytically~\cite{Arbuzov:2002pp,Arbuzov:2002cn}. A few years
later, the full spectrum was calculated with a numerical loop
integration~\cite{Anastasiou2005The-electron} and the original
calculation of \cite{vanRitbergen:1999fi} was extended to include mass
effects~\cite{Pak:2008qt}.  It was only recently that the form factors
necessary for a fully differential calculation were
published~\cite{Chen:2018dpt,Engel:2018fsb}.

We have included muon and electron loops but neither tau nor hadronic
contributions~\cite{vanRitbergen:1998hn,Davydychev:2000ee}. We treat
the electromagnetic coupling $\alpha$ in the on-shell scheme, except in
Table~\ref{tab:stuart} where, in order to compare
to~\cite{vanRitbergen:1999fi} we need the $\MS$ coupling
$\bar\alpha \equiv \bar\alpha(\mu=M)$.

%%%%%%%%%%%%%%%%%%%%
\subsection{Calculation}
%%%%%%%%%%%%%%%%%%%%

The momenta of the muon and electron are written as
\begin{align}
p_1 = M(1,\vec 0,0)\,,
\qquad
p_2 = \frac{x_e}{2}\, M(1,\vec 0,\beta)\,,
\end{align}
where $x=(p_1-p_2)^2/M^2$ and
\begin{align}
z=\frac{m}{M}\equiv \frac{m_e}{m_\mu} \,,
\qquad
x_e \equiv 1-x+z^2 \, .
\end{align}
Furthermore,
$\beta$ is the velocity of the electron in the muon rest frame.

Apart from the form factors needed for $\bbit{2}{n}$, we also need
matrix elements for $\bbit{2}{n+1}$ and $\bbit{2}{n+2}$.\footnote{Note
  that, to remain consistent with the discussion above, we will denote
  the decay by $\sigma$ instead of $\Gamma$.}  We have generated the
diagrams for $\M{n+1}1$ and $\M{n+2}0$ using
QGraf~\cite{Nogueira:1991ex} and calculated them using
Package-X~\cite{Patel:2015tea}. The numerical integration of
$\bbit{2}{f}$ and $\bbit{2}{hh}$ was performed in {\tt Fortran} using
{\tt vegas}~\cite{Lepage:1980jk}. Most loop integrals in $\M{n+1}1$
were included explicitly, while the more complicated triangle- and
box-functions were evaluated using the {\tt COLLIER} library~\cite{
  Denner:2016kdg}.

%%%%%%%%%%%%%%%%%%%%
\subsection{\texorpdfstring{$\xc$}{xicut} dependence} \label{sec:xic}
%%%%%%%%%%%%%%%%%%%%

Due to the simplicity of the process, it is actually possible to
explicitly compute the contribution of the integral
$\mathcal{I}(\xi_{c_A},\xi_{c_B})$ of \eqref{eq:nnlo:sin} to the total
decay rate, and check that the dependence of all four $\xc$ parameters
vanishes for the physical result. To verify this we perform weighted
two-dimensional fits of the form $c_{11} \log\xi_{c_i}\log\xi_{c_j} +
c_{10} \log\xi_{c_i}+c_{01} \log\xi_{c_j}+c_{00}$ for the numerical
data of $\bbit{2}{f}$ and $\bbit{2}{hh}$ and check whether the $\xi_c$
dependences vanish within the numerical error of this fit.

\begin{figure}
\centering
\subfigure[$\xi_{c_A}=0.003$]{
\includegraphics[width=0.8\textwidth]{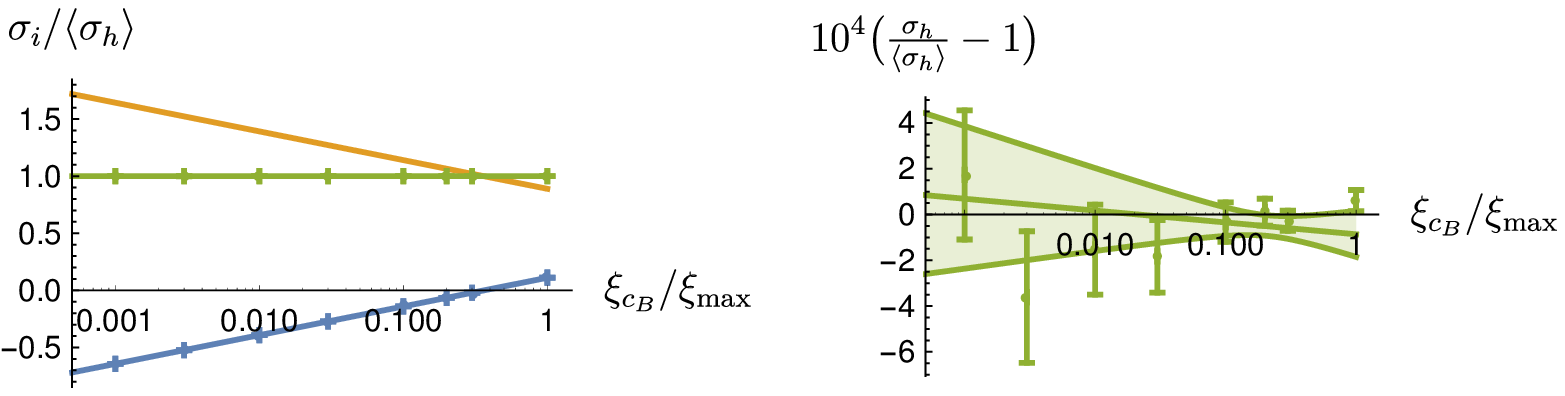}
}
\subfigure[$\xi_{c_A}=0.03$]{
\includegraphics[width=0.8\textwidth]{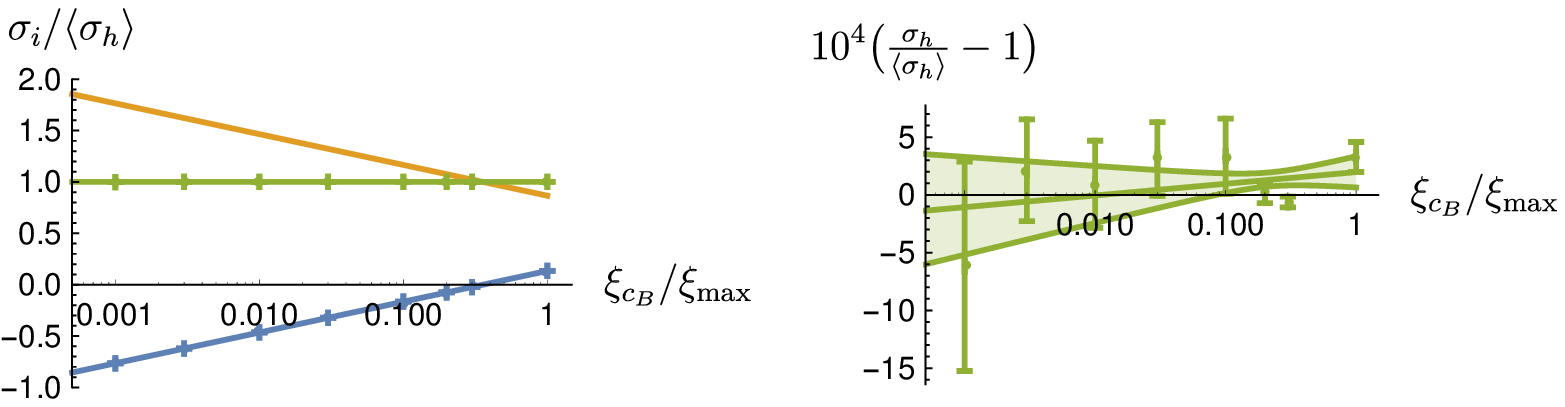}
}
\caption{$\xi_{c_B}$ (in)dependence of the ($\epsilon^0$ coefficient of the)
  real-virtual contribution, for two values of $\xi_{c_A}$. The blue
  dots show the $\bbit{2}{f}$ contribution with a fit (blue line).
  The orange line corresponds to $\bbit{2}{d} =
  -\mathcal{I}(\xi_{c_A},\xi_{c_B})$. In green we show the sum
  $\bbit{2}{h}$, \eqref{eq:nnloh}. The right panel is zoomed into the
  region of interest showing $1\sigma$ confidence bounds for the
  fit. All plots are normalised such that the average $\langle
  \bbit{2}{h}(\xi_{c_A})\rangle=1$.
}

\label{fig:xib}
\end{figure}

\begin{figure}
\centering
\includegraphics[width=0.8\textwidth]{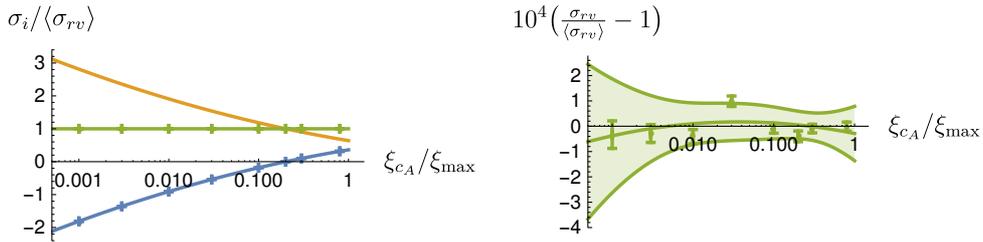}
\caption{$\xi_{c_A}$ (in)dependence of the ($\epsilon^0$ part of the)
  real-virtual contribution. The combined hard contributions
  $\bbit{2}{h}$ (blue, dots and fit) and the soft contribution
  $\bbit{2}{s}$ (orange) sum up to the total real-virtual contribution
  $\bbit{2}{rv}$ (green), \eqref{eq:rv}. The right panel is zoomed
  into the region of interest, showing $1\sigma$ confidence bounds to
  the fit.  All plots are normalised such that $\langle
  \bbit{2}{rv}\rangle=1$.
}
\label{fig:xia}
\end{figure}

The $\xi_{c_B}$ (in)dependence (according to \eqref{eq:nnloh}) of the
combination $\bbit{2}{h} = \bbit{2}{f}-\mathcal{I}$ is shown in
Figure~\ref{fig:xib} for two different values of $\xi_{c_A}$.  To
numerically evaluate these expressions in the plots, we drop the
$1/\epsilon$ poles consistently in all intermediate expressions.  In a
next step, $\bbit{2}{h}$ is then combined with $\bbit{2}{s}$ in
Figure~\ref{fig:xia}. As indicated in \eqref{eq:rv} this has to result
in a $\xi_{c_A}$ independent expression.

\begin{figure}
\centering
\subfigure[$\xi_{c_1}=0.001$]{
\includegraphics[width=0.8\textwidth]{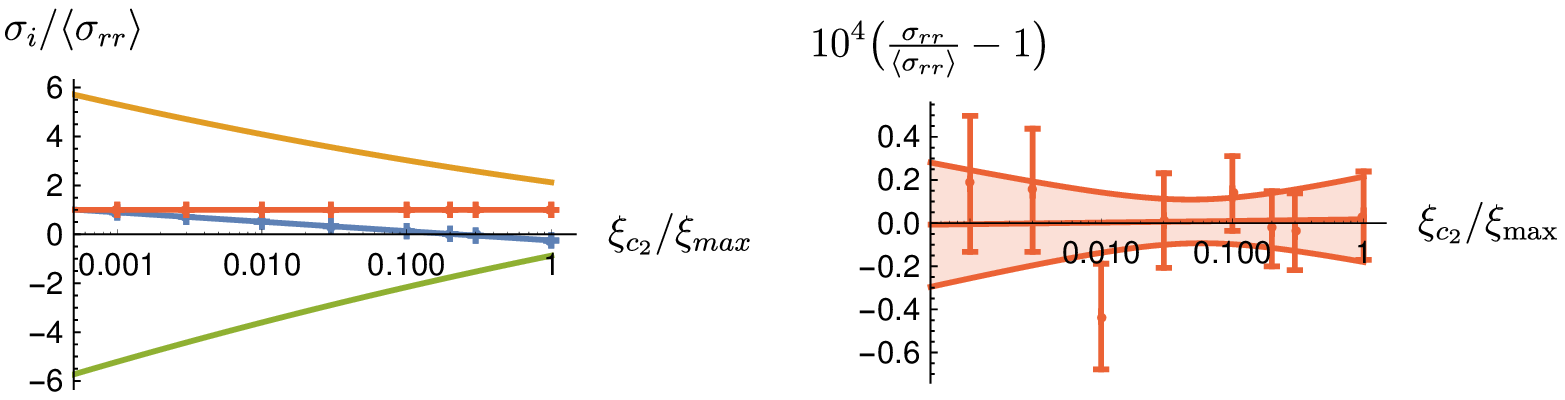}
}
\subfigure[$\xi_{c_1}=0.3$]{
\includegraphics[width=0.8\textwidth]{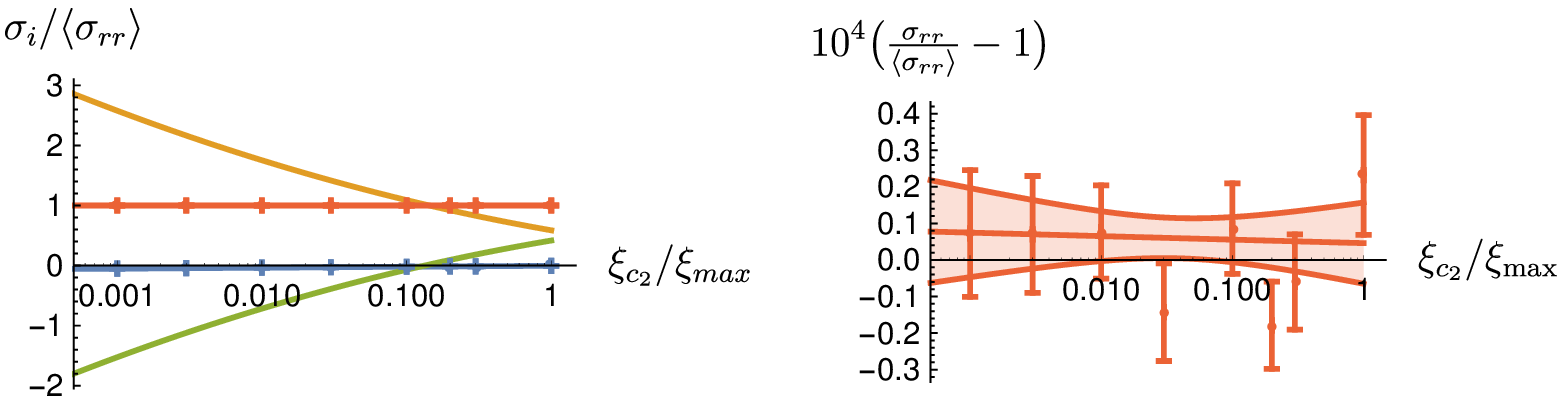}
}

\caption{The $\xi_{c_2}$ (in)dependence of the ($\epsilon^0$
  coefficient of the) double-real corrections for two $\xi_{c_1}$. The
  $\bbit{2}{hh}$-data (blue, data and fit), $\bbit{2}{ss}$ (yellow)
  and $\tfrac12(\mathcal{I}(\xi_{c_1},\xi_{c_2}) +
  \mathcal{I}(\xi_{c_1},\xi_{c_2}))$ contribution (green) sum up to
  the total $n+2$ particle contributions (red). The right panel
  magnifies the region of interest including $1\sigma$ bounds. Note
  that the swap $c_1\leftrightarrow c_2$ is trivial. All plots are
  normalised such that $\langle \bbit{2}{rr}(\xi_{c_2})\rangle=1$.
}
\label{fig:xir}
\end{figure}

Similarly, we have shown in Figure~\ref{fig:xir} the $\xi_{c_2}$
(in)dependence of the double-real corrections $\bbit{2}{rr}$
(cf. \eqref{eq:rr}) for some examples of $\xi_{c_1}$.

\begin{figure}
\centering

\begin{tabular}{l|ccc|c}

          & $\sigma_\gamma^{(2)}/\sigma_0$ &$\sigma_\mu^{(2)}/\sigma_0$ &$\sigma_e^{(2)}/\sigma_0$ & total\\\hline
massified & 3.42    & -0.0364 & 3.24    & 6.62 \\
massive   & 3.54    & -0.0364 & 3.16    & 6.66 \\
\hline
massless~\cite{vanRitbergen:1999fi}    & 3.56    & -0.0364 & 3.22    & 6.74 \\
\hline
 $\Delta_{\text{rel}}$ massified & $3.7\times10^{-2}$ & $ 0$
& $6.1\times 10^{-3}$ & $1.6\times10^{-2}$ \\
 $\Delta_{\text{rel}}$ massive      & $5.0\times10^{-3}$ &
$1.9\times10^{- 4}$
& $2.0\times 10^{-2}$ & $1.1\times10^{-2}$ \\
\end{tabular}

\renewcommand{\figurename}{Table}
\caption{The different contributions to $\sigma^{(2)}$. Note that
  $\sigma_e^{(2)}$ also includes the process $\mu\to\nu\bar\nu e\,
  ee$. See text for interpretation. The coupling $\bar\alpha(\mu=M)$
  is renormalised in the $\MS$ scheme. $\Delta_{\text{rel}}$ denotes
  the relative difference of our results to the massless result
  \cite{vanRitbergen:1999fi}. }
\label{tab:stuart}
\end{figure}

%%%%%%%%%%%%%%%%%%%%
\subsection{Results for the decay rate}
%%%%%%%%%%%%%%%%%%%%

The first quantity we consider is the full decay width
\begin{align}
\label{def:resnnlo}
\sigma_2 = \sigma_0  + \frac{\bar\alpha}{\pi}\, \sigma^{(1)}  
+ \Big(\frac{\bar\alpha}{\pi}\Big)^2\, \sigma^{(2)} 
+ \mathcal{O}(\bar\alpha^3) \, ,
\end{align}
where we have pulled out factors of the $\MS$ coupling
$\bar\alpha/\pi$.  We compute $\sigma^{(2)}$ using the massified form
factors as well as the form factor with full $m$
dependence~\cite{Engel:2018fsb, Chen:2018dpt}. We will label these two
results `massified' and `massive', respectively.  In the case of the
massified result, we expand all three parts of the integrand
contributing to $\sigma^{(2)}_n$, see \eqref{eq:nnloind0} and
\eqref{eq:nnlo:n}. Of course, the exact mass dependence of
$\bbit{1}{s}$ and $\bbit{2}{ss}$ is usually much easier to obtain than
for $\M n2$. However, the complete cancellation of singularities requires
a consistent expansion in $z$ of all contributions at the $n$-particle
level.

Because the full decay rate does not contain terms $\log z \sim \log
m$ the limit $m\to 0$ exists and we can compare our massified and
massive results with the result for a massless
electron~\cite{vanRitbergen:1999fi}.  We note that in this particular
case (contrary to distributions, where $\log m$ terms exist), the
massified result is not expected to be superior to the massless
computation.

Following \cite{vanRitbergen:1999fi}, we split the result into three
parts: photonic corrections $\sigma_\gamma^{(1)}$ and
$\sigma_\gamma^{(2)}$, corrections due to an electron pair (real or
virtual) $\sigma_e^{(2)}$, and corrections due to a muon pair
(virtual) $\sigma_\mu^{(2)}$. These parts have been defined and their
analytic results in the massless case given in equations (2.11),
(2.13) and (2.15) of \cite{vanRitbergen:1999fi}.  The individual
results for the NNLO corrections are shown in Table~\ref{tab:stuart},
where the Monte Carlo error is smaller than the significant
digits. Note that~\cite{vanRitbergen:1999fi} had to include the
`open-lepton production' $\mu\to\nu\bar\nu e\, ee$ into their
calculation of $\sigma_e^{(2)}$ to guarantee finiteness.  We have
included this process as well~\cite{Pruna:2016spf} since it
contributes to $\sigma_e^{(2)}$ (two-trace contribution) and
$\sigma_\gamma^{(2)}$ (one-trace contribution).\footnote{The amplitude
  for $\mu^-\to\nu\bar\nu e^-\,e^+e^-$ has a (anti)symmetry under
  exchange of the two $e^-$.  This gives rise to two types of
  interference terms in the matrix element: first the contribution
  that is also present without this symmetry (two-trace) and one where
  the swapped is interfered with the non-swapped contribution
  (one-trace).}

The results of Table~\ref{tab:stuart} merit a few comments:
\begin{itemize}

    \item
    The good agreement for the purely photonic contributions
    $\sigma_\gamma^{(2)}$ between the massive and massless result is
    due to the absence of terms $\log m$ and $m\log m$ as discussed
    by~\cite{vanRitbergen:1999fi}.

    \item
    The massified results differs by about 3\% from the massive (and
    massless) result for $\sigma_\gamma^{(2)}$. This is due to the
    mismatch between the real corrections, that were calculated with
    the full electron mass dependence, and the massified two-loop
    amplitude that only includes logarithmically enhanced mass
    effects. 

    \item
    The massified results agrees perfectly
    with~\cite{vanRitbergen:1999fi} for the $\sigma_\mu^{(2)}$ part
    because the contribution comes purely from one two-loop diagram
    that is free of any soft or collinear logarithms and hence
    effectively massless.

    \item
    The massive and massless results for $\sigma_e^{(2)}$ agree only
    up to two percent. This difference can be accounted for through
    the two-trace contribution of the open-lepton production. In the
    pure electron trace $m$ must not be neglected to lead to finite
    expressions. However, in the other trace the electron mass can be
    set to zero.  Our value of $3.16$ was calculated with full
    electron mass dependence. If we were to set $m\to0$ in the this
    trace, we would obtain $3.23$ in much better agreement
    with~\cite{vanRitbergen:1999fi}.

    \item
    The $\sigma_e^{(2)}$ part contains the factorisation anomaly,
    already discussed in~\cite{Engel:2018fsb}.

\end{itemize}
Note that in any case the `massive' result should be considered the
reference. Our results agree with~\cite{Pak:2008qt}.  For the pure
mass effects of the photonic part, this agreement is only at the 20\%
level. This is due to large numerical cancellations between
$\sigma^{(2)}_n$, $\sigma^{(2)}_{n+1}$ and $\sigma^{(2)}_{n+2}$ which
make the extraction of a few-percent effect on the NNLO corrections
numerically challenging. In fact, an efficient numerical evaluation of
the integrals with full mass dependence~\cite{Chen:2018dpt} has only
recently been implemented~\cite{Naterop:2019}.

%%%%%%%%%%%%%%%%%%%%
\subsection{The electron energy spectrum}
%%%%%%%%%%%%%%%%%%%%

\begin{figure}
\centering
\includegraphics[width=0.8\textwidth]{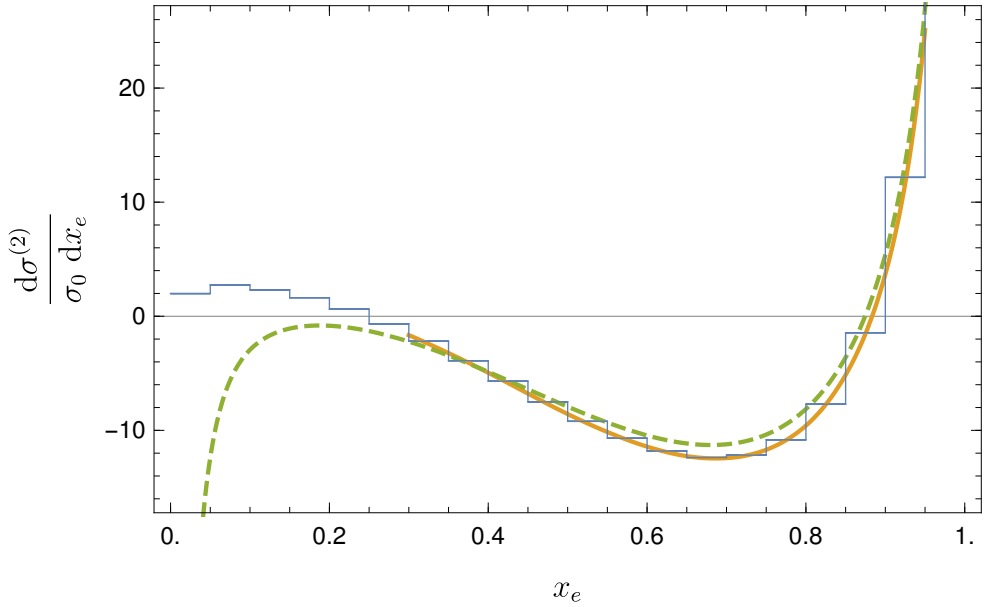}
\caption{The NNLO corrections to the  electron energy
  spectrum, omitting a factor $(\alpha/\pi)^2$. The logarithmic
  contributions of~\cite{Arbuzov:2002pp,Arbuzov:2002cn} (green dashed)
  agree reasonably well with our massive result (blue histogram) for large
  $x_e$. Adding the constant terms
  of~\cite{Anastasiou2005The-electron} (orange) we obtain good
  agreement. }
\label{fig:xe}
\end{figure}

In order to validate our computation, we consider the NNLO corrections
to the normalised electron energy spectrum $x_e=2E/M$ and compare them
to results available in the literature. If two (negatively charged)
electrons are present in the final state, we include both of them in
the $x_e$ distribution.  The leading and sub-leading logarithmic
contributions for this observable were calculated
in~\cite{Arbuzov:2002pp,Arbuzov:2002cn}.  Because this corresponds to
a strict expansion in $z$, we expect good agreement for large $x_e$ as
noticed in~\cite{Engel:2018fsb}.  In Figure~\ref{fig:xe} we compare
the two results and see that the differences are compatible with the
constant (logarithm-free) terms missing
in~\cite{Arbuzov:2002pp,Arbuzov:2002cn}. These terms were computed
numerically and shown in a plot for $x_e>0.3$
in~\cite{Anastasiou2005The-electron}. If we include these constant
terms of~\cite{Anastasiou2005The-electron}, we obtain perfect
agreement with our result, using the massive form factors.  Note that
the difference between massified and massive result in
Figure~\ref{fig:xe} is at the percent level and only becomes visible
around the zero crossing at $x_e\approx0.21$ and $x_e\approx0.88$,
never changing the overall picture. The on-shell coupling
$(\alpha/\pi)^2$ is omitted in the results shown in the
Figure~\ref{fig:xe}.

With a fully differential Monte Carlo code, we can compute arbitrary
distributions, including cuts. As an example, we consider again the
normalised electron energy spectrum but impose a cut on photon
emission. Concretely, we restrict the total energy of all photons
within a cone of angle $\theta\equiv\sphericalangle(\vec p_e, \vec
p_\gamma) = 37^\circ$ (i.e. a cone with $|\cos\theta|>0.8$) around the
electron to be less than 10~MeV.

\begin{figure}
\centering
\includegraphics[width=0.7\textwidth]{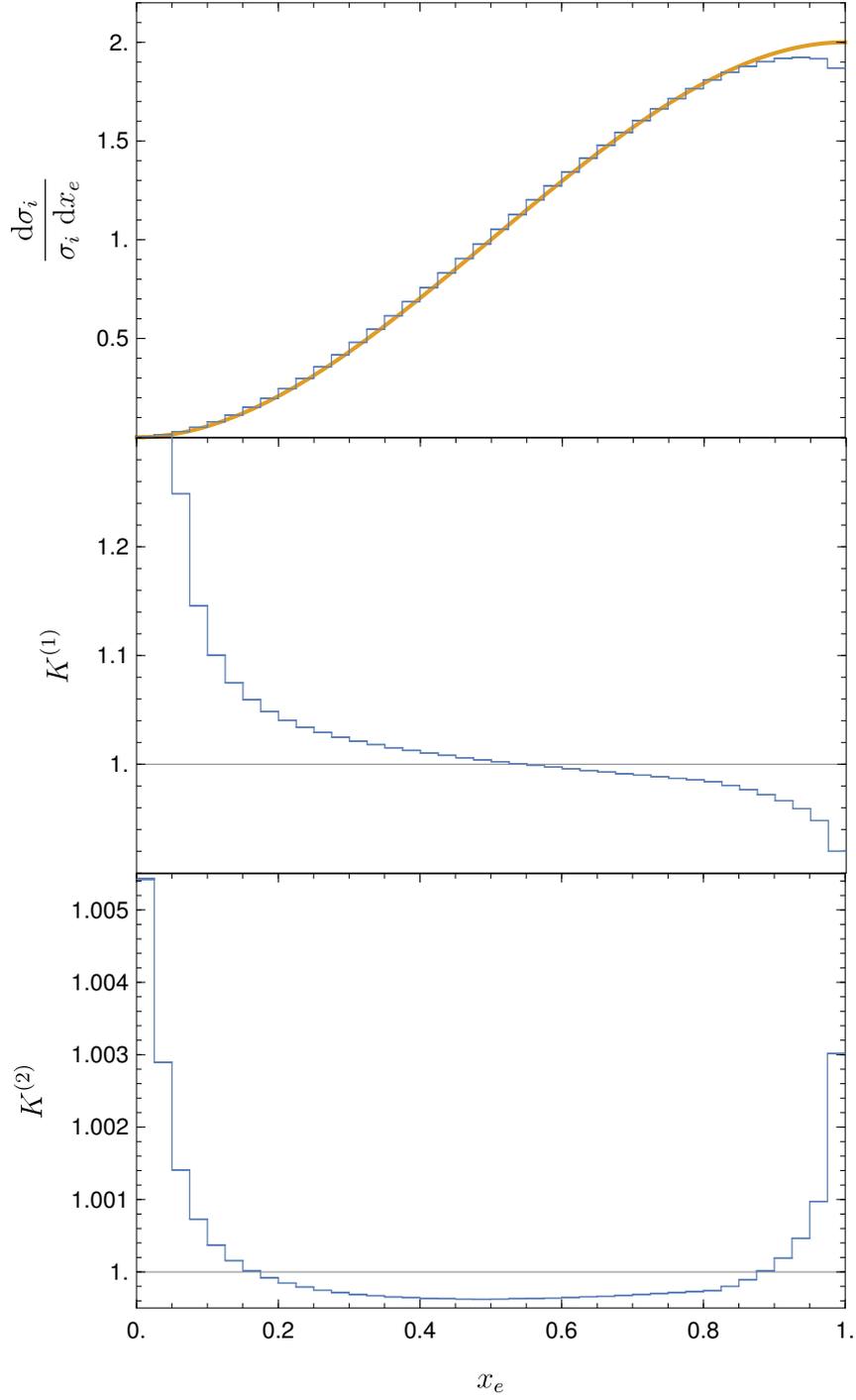}
\caption{Top panel: The normalised electron energy spectrum at LO 
  (orange line) and NNLO (blue histogram) with a cut on photon
  emission. The middle and bottom panel show the NLO and NNLO $K$
  factors, respectively.}
\label{fig:xecut}
\end{figure}

The results are shown in Figure~\ref{fig:xecut}. Comparing the
normalised NNLO result (blue histogram) to the normalised LO result
(orange line) in the top panel reveals that only for large $x_e$ the
corrections to the shape are relevant. This is driven by the NLO
corrections. They are large at both ends of the $x_e$ spectrum, as
shown by the NLO $K^{(1)}$ factor 
\begin{align}
K^{(i)} = 1 + \Big(\frac{\alpha}{\pi}\Big)^i\
   \frac{\D\sigma^{(i)}/\D x_e}{\D\sigma_0/\D x_e}
\end{align}
in the middle panel. Typically, the NNLO corrections (shown in the
bottom panel) are below 0.1~\% and even in the regions of huge NLO
corrections they are below 0.5\%.

%%%%%%%%%%%%%%%%%%%%%%%%%%%%%%%%%%%%%%%%%%%%%%%%%%%%%%%%%%%%%%%%%%%%%
\section{Conclusion}\label{sec:conclusion}
%%%%%%%%%%%%%%%%%%%%%%%%%%%%%%%%%%%%%%%%%%%%%%%%%%%%%%%%%%%%%%%%%%%%%

We have presented a subtraction scheme tailored to the case of QED
with massive fermions. This allows to perform the phase-space
integration and obtain predictions for arbitrary physical cross
sections if the corresponding matrix elements are known. While we have
primarily NNLO calculations in mind, the extension of the scheme
beyond NNLO is also discussed. After describing its implementation, we
have commented on its properties. We have noted that, while intermediary
results may look complicated, most of this complexity drops out in the
final result. We have verified our scheme by calculating the muon
decay at NNLO and compared with known results.

While our scheme leads to very simple expressions, there are still
several avenues for further developments. Since the analytic
computation of matrix elements with massive fermions is very
challenging, one possibility is to investigate the option of computing
the subtracted matrix elements numerically. After all, these are
finite expressions. If it is possible to perform the subtraction and
UV renormalisation at the integrand level, a direct numerical
evaluation in four dimensions should be feasible.

Even if the matrix elements are available, the numerical integration
can be challenging. First of all, the matrix elements need to be
implemented in a stable way in all corners of phase space. The
presence of small fermion masses also results in pseudo-collinear
singularities that can lead to numerical instabilities. An option to
deal with these is to treat them as singularities, i.e. subtract them
and add them back in integrated form. Obviously, this tarnishes the
simplicity of the subtraction scheme. But it offers a possibility to
take into account small masses in an expanded form for the
double-real and real-virtual corrections. Such an implementation can
be seen as a generalisation of massification for all parts of a NNLO
cross-section computation.

The scheme we have presented relies on eikonal subtraction and is
closely related to the YFS exponentiation of the soft singularities. The
YFS picture has been used to construct Monte Carlo codes that describe
multiple emission of soft photons. Thus, our scheme for fixed-order
computations lends itself to be combined with a YFS Monte Carlo
program. Such a generalisation will allow to resum large logarithms
and combine this with fixed order results. Ultimately, this is
required to obtain very precise predictions for fully differential QED
observables.

%%%%%%%%%%%%%%%%%%%%%%%%%%%%%%%%%%%%%%%%%%%%%%%%%%%%%%%%%%%%%%%%%%%%%
\subsection*{Acknowledgements}
%%%%%%%%%%%%%%%%%%%%%%%%%%%%%%%%%%%%%%%%%%%%%%%%%%%%%%%%%%%%%%%%%%%%%

The authors would like to thank Pulak Banerjee for commenting on early
versions of the subtraction scheme and the draft. We also would like
to thank him for his contributions to improving the evaluation of
$\fM{n+1}1$. Additionally we would like to thank Dario M\"{u}ller and
Fiona Kirk for commenting on the readability of the manuscript.

This scheme was presented at the `\nth{2} Workstop / Thinkstart --
theory for muon-electron scattering @ 10ppm' that was held at the
University of Zurich between the \nth{4} and \nth{7} of February 2019.
We would like to thank all participants for their comments.

TE and YU acknowledge support by the Swiss National Science Foundation
(SNF) under contract 200021\_178967.

\appendix

%%%%%%%%%%%%%%%%%%%%
\section{The integrated eikonal \texorpdfstring{$\ieik$}{Ehat}} \label{app:eikonal}
%%%%%%%%%%%%%%%%%%%%

Here we give in our conventions the explicit form of integrated
eikonal required for massive QED. These expressions have been computed
in \cite{Frederix2009Automation}. As discussed in the text, we do not
need terms $\mathcal{O}(\epsilon)$ or higher.

We start with defining a few auxiliary quantities:
\begin{align*}
\beta_j &\equiv \sqrt{1-\frac{m_j^2}{E_j^2}}\ , &
v_{kj} &\equiv \sqrt{1-\left(\frac{m_j m_k}{p_j\cdot p_k}\right)^2 }\ , &%\\
a_{kj} &\equiv (1+v_{kj})\frac{p_j\cdot p_k}{m_k^2} \ , &
%\lambda_{kj} &\equiv a_{kj}\, E_k - E_j\ ; &
\nu_{kj} &\equiv \frac{a_{kj}^2 m_k^2 - m_j^2}{2 \, (a_{kj}\, E_k - E_j)}\ . &
\end{align*}
Following \cite{Frederix2009Automation}, the integrated eikonal can
then be written as
\begin{align}
\label{app:ieik}
\ieik_{kj} = \frac{\alpha}{2\pi} 
\frac{(4\pi)^\epsilon}{\Gamma(1-\epsilon)}
\left(\frac{\xc^2\, s}{\mu^2}\right)^{-\epsilon} 
\bigg(&-\frac{1}{2\epsilon} \frac{1}{v_{kj}}
\log\frac{1+v_{kj}}{1-v_{kj}} 
\\ \nonumber 
& +
 \frac{a_{kj} (p_j\cdot p_k) }{2\, (a_{kj}^2 m_k^2 - m_j^2)}
\Big( J(a_{kj} E_k, \beta_k,\nu_{kj}) - J(E_j,\beta_j,\nu_{kj}) \Big)
\bigg) + \mathcal{O}(\epsilon)\, ,
\end{align}
where we have used the function
\begin{align*}
J(x,y,z) &= 
\Bigg( \log^2\frac{1-y}{1+y} 
+ 4\, \text{Li}_2\left(1-\frac{x(1+y)}{z}\right)
+ 4\, \text{Li}_2\left(1-\frac{x(1-y)}{z}\right)
\Bigg)
\end{align*}
For the case $j=k$ this expression simplifies to 
\begin{align}
\label{app:ieikk}
\ieik_{jj} &= \frac{\alpha}{2\pi} 
\frac{(4\pi)^\epsilon}{\Gamma(1-\epsilon)}
\left(\frac{\xc^2\, s}{\mu^2}\right)^{-\epsilon} 
\bigg(-\frac{1}{\epsilon} - \frac{1}{\beta_j}
  \log\frac{1+\beta_j}{1-\beta_j} \bigg) + \mathcal{O}(\epsilon)\,
\end{align}
for the self-eikonals.

%%%%%%%%%%%%%%%%%%%%%%%%%%%%%%%%%%%%%%%%%%%%%%%%%%%%%%%%%%%%%%%%%%%%%
\section{Details for \texorpdfstring{FKS$^3$}{FKS3}} \label{app:nnnlo}
%%%%%%%%%%%%%%%%%%%%%%%%%%%%%%%%%%%%%%%%%%%%%%%%%%%%%%%%%%%%%%%%%%%%%

%%%%%%%%%%%%%%%%%%%%
\subsection{Real-virtual-virtual contribution}
%%%%%%%%%%%%%%%%%%%%

Let us begin with the real-virtual-virtual part that we split again
into a hard and soft contribution
\begin{align}
\label{eq:nnnlorvv}
\bbit{3}{rvv} = \D\Phi_{n+1} \M{n+1}2 = \bbit{3}{s}(\xc) + \bbit{3}{h}(\xc)
\end{align}
as in \eqref{eq:shnnlo}.  Using that even at the two-loop level
\begin{align*}
\mathcal{S}_{n+1}\M{n+1}2 = \eik_{n+1}\M{n}2\,,
\end{align*}
the soft contribution in analogy to \eqref{eq:nlo:s}
and \eqref{eq:nnlo:s}  is given by
\begin{align*}
\bbit{3}{s}(\xc) \to \D \Phi_{n}\ieik(\xc) \M{n}2\,.
\end{align*}
The hard contribution  is now
\begin{align*}
\bbit{3}{h}(\xc) &= 
  \pref1 \D\xi\, \cdis{\xi^{1+2\epsilon}} \big(\xi^2\M{n+1}2\big)
\\&
 =\pref1 \D\xi\, \cdis{\xi^{1+2\epsilon}} \xi^2\Big(
     \fM{n+1}2 -\ieik(\xc)\M{n+1}1-\frac1{2!}\ieik(\xc)^2\M{n+1}0
\Big)
\\&
 =\bbit{3}{f}(\xc) + \underbrace{\bbit{3}{d1}(\xc) + \bbit{3}{d0}(\xc)}_{\bbit{3}{d}(\xc)}
\end{align*}
where $\bbit{3}{f}$ is finite and the divergent part $\bbit{3}{d}$
is composed of
\begin{align*}
\int \bbit{3}{d1}(\xc) &=
-
\int  \pref1 \D\xi\, \cdis{\xi^{1+2\epsilon}} \xi^2\Big(
      \ieik(\xc)\M{n+1}1
\Big) \equiv -\mathcal{I}^{(1)}(\xc)\,,\\
\int  \bbit{3}{d0}(\xc) &=
- \int \frac1{2!} \pref1 \D\xi\, \cdis{\xi^{1+2\epsilon}} \xi^2\Big(
      \ieik(\xc)^2\M{n+1}0
\Big) \equiv -\frac1{2!}\mathcal{J}(\xc)\,.
\end{align*}
Above we have defined two functions $\mathcal{I}^{(1)}$ and
$\mathcal{J}$ that are potentially tedious to compute. However, as we
will see they cancel in the final result, similar to the function
$\mathcal{I}$ at NNLO.

%%%%%%%%%%%%%%%%%%%%
\subsection{Real-real-virtual contribution}
%%%%%%%%%%%%%%%%%%%%

The real-real-virtual contribution are similar to the double-real
contribution of FKS$^2$
\begin{align}
& \qquad
\bbit{3}{rrv} = \D\Phi_{n+2} \M{n+2}{1} = \bbit{3}{ss}(\xc) +
\bbit{3}{sh}(\xc) + \bbit{3}{hs}(\xc) + \bbit{3}{hh}(\xc)
\,,\\[5pt] \nonumber
& \left\{\begin{array}{c}
   \bbit{3}{ss}(\xc) \\[5pt] 
   \bbit{3}{hs}(\xc) \\[5pt]  
   \bbit{3}{sh}(\xc) \\[5pt]  
   \bbit{3}{hh}(\xc)
\end{array}\right\} =
    \pref2\ \frac1{2!}\ 
\left\{\begin{array}{c}
   \frac{\xc^{-2\epsilon}}{2\epsilon}\delta(\xi_1) \,
   \frac{\xc^{-2\epsilon}}{2\epsilon}\delta(\xi_2)
   \\[3pt] 
  -\frac{\xc^{-2\epsilon}}{2\epsilon}\delta(\xi_2) \,
   \cdis{\xi_1^{1+2\epsilon}}
   \\[3pt]
  -\frac{\xc^{-2\epsilon}}{2\epsilon}\delta(\xi_1) \,
   \cdis{\xi_2^{1+2\epsilon}}
   \\[3pt]
   \cdis{\xi_1^{1+2\epsilon}}\,
   \cdis{\xi_2^{1+2\epsilon}}
\end{array}\right\}\, \D\xi_1\,\D\xi_2
\ \xi_1^2\xi_2^2\M{n+2}1\,.
\end{align}
Obviously $\int\bbit{3}{hs}=\int\bbit{3}{sh}$ and
\begin{align}
\int\bbit{3}{hs}(\xc) &=
    \pref2 \ \frac1{2!}\ \int\D\xi_1\ \cdis{\xi_1^{1+2\epsilon}}
    \big(\xi_1^2\M{n+1}1\big)
    \ieik(\xc)
= \frac1{2!}\,\mathcal{I}^{(1)}(\xc) \, .
\end{align}
Furthermore, as for \eqref{eq:nnlo:ss} we find 
\begin{align}
\bbit{3}{ss}(\xc) &\to\ \D\Phi_{n} \frac1{2!} \ieik(\xc)^2 \M{n}1 \,.
\end{align}
The hard contribution is  not yet finite due to the
explicit $1/\epsilon$ pole in $\M{n+2}1$. As is customary by now we
again perform an eikonal subtraction 
\begin{align*}
    \M{n+2}1 \equiv \fM{n+2}1(\xc) - \ieik(\xc)\,\M{n+2}0\,.
\end{align*}
and write
\begin{align*}
\bbit{3}{hh}(\xc) &= \bbit{3}{hf}(\xc) + \bbit{3}{hd}(\xc)\,,\\
\bbit{3}{hf}(\xc) &= \pref2\ \frac1{2!}\ 
   \cdis{\xi_1^{1+2\epsilon}}\,
   \cdis{\xi_2^{1+2\epsilon}}\,
   \xi_1^2\xi_2^2\fM{n+2}1(\xc)\,,\\
\int\bbit{3}{hd}(\xc) &=-\int \pref2\ \frac1{2!}\ 
   \cdis{\xi_1^{1+2\epsilon}}\,
   \cdis{\xi_2^{1+2\epsilon}}\,
   \xi_1^2\xi_2^2\ieik(\xc)\,\M{n+2}0 \equiv -\frac1{2!} \mathcal{K}(\xc)\,.
\end{align*}
Here we have defined a third auxiliary function $\mathcal{K}$ that
will cancel in the final result. 

%%%%%%%%%%%%%%%%%%%%
\subsection{Triple-real contributions}
%%%%%%%%%%%%%%%%%%%%

The evaluation of the triple-real contributions proceeds along the
lines of the FKS$^2$ double-real part, albeit with more (individually
$\xc$ dependent) terms
\begin{align*}
\bbit{3}{rrr} &= \D\Phi_{n+3} \M{n+3}0 = 
  \bbit{3}{hhh} 
+ \underbrace{\bbit{3}{hhs} + \bbit{3}{hsh} + \bbit{3}{shh}}_{3\bbit{3}{hhs}}
+ \underbrace{\bbit{3}{hss} + \bbit{3}{shs} + \bbit{3}{ssh}}_{3\bbit{3}{hss}}
+ \bbit{3}{sss}\,.
\end{align*}
Because we choose all $\xc$ equal, it does not matter which photon is
soft, just how many. Thus, we are left with four different kinds of
contributions
\begin{align*}
\left\{\begin{array}{c}
   \bbit{3}{sss}(\xc) \\[5pt]  
   \bbit{3}{hss}(\xc) \\[5pt]  
   \bbit{3}{hhs}(\xc) \\[5pt]  
   \bbit{3}{hhh}(\xc)
\end{array}\right\} &=
%    \pref3\ 
   \prod_{i=1}^3 \Big( \D\Upsilon_i \D\xi_i\, \xi_i^2\Big)\ \D \Phi_{n,3}
   \frac{\M{n+3}0}{3!}\ 
\left\{\begin{array}{c}
   \frac{\xc^{-2\epsilon}}{2\epsilon}\delta(\xi_1) \,
   \frac{\xc^{-2\epsilon}}{2\epsilon}\delta(\xi_2) \,
   \frac{\xc^{-2\epsilon}}{2\epsilon}\delta(\xi_3)
   \\[3pt]
   \frac{\xc^{-2\epsilon}}{2\epsilon}\delta(\xi_2) \,
   \frac{\xc^{-2\epsilon}}{2\epsilon}\delta(\xi_3) \,
   \cdis{\xi_1^{1+2\epsilon}}
   \\[3pt]
  -\frac{\xc^{-2\epsilon}}{2\epsilon}\delta(\xi_3) \,
   \cdis{\xi_1^{1+2\epsilon}}
   \cdis{\xi_2^{1+2\epsilon}}
   \\[3pt]
   \cdis{\xi_1^{1+2\epsilon}}\,
   \cdis{\xi_2^{1+2\epsilon}}\,
   \cdis{\xi_3^{1+2\epsilon}}
\end{array}\right\}\, .
\end{align*}
The triple-hard $\bbit{3}{hhh}$ contribution is finite and can be
integrated numerically. For the triple-soft $\bbit{3}{sss}$ we get
\begin{align*}
\bbit{3}{sss}(\xc) &= \D\Phi_n\,  \frac1{3!} \ieik^3\,\M n0\,.
\end{align*}
The double-soft contribution can be expressed in terms of the function
$\mathcal{J}(\xc)$ as
\begin{align*}
\int \bbit{3}{hss}(\xc) &=\int 
    \pref1\ \frac1{3!}\ 
   \ieik(\xc)^2\cdis{\xi_1^{1+2\epsilon}}
   \D\xi_1 \ \xi_1^2\M{n+1}0 = \frac1{3!} \mathcal{J}(\xc)\,.
\end{align*}
Similarly, the single-soft contribution
\begin{align*}
  \int  \bbit{3}{hhs}(\xc) &= \int \pref2\ \frac1{3!}\ \ieik(\xc)
   \cdis{\xi_1^{1+2\epsilon}} \cdis{\xi_2^{1+2\epsilon}}\,
   \D\xi_1\,\D\xi_2 \ \xi_1^2\xi_2^2\M{n+2}0 =
   \frac1{3!}\mathcal{K}(\xc)\,.
\end{align*}
involves the auxiliary function  $\mathcal{K}$.

%%%%%%%%%%%%%%%%%%%%
\subsection{Combination}
%%%%%%%%%%%%%%%%%%%%

Combining all contributions at N$^3$LO we need to evaluate
\eqref{eq:nnnlocomb}. Collecting the terms with an $n$-parton phase
space we get 
\begin{align*}
\bbit{3}{n}(\xc)  &= \int\D\Phi_n\Big(
     \M{n}3
 + \underbrace{\ieik(\xc) \M{n}2}_{\bbit{3}{s}}
 + \underbrace{\frac1{2!} \ieik(\xc)^2 \M{n}1 }_{\bbit{3}{ss}}
 + \underbrace{1\times\frac1{3!}\ieik(\xc)^3\M{n}0}_{\bbit{3}{sss}} \Big)
\\&\qquad
  \underbrace{ -\mathcal{I}(\xc)
    - \frac1{2!} \mathcal{J}(\xc)}_{\bbit{3}{d}}
 + \underbrace{\frac1{2!}
   \mathcal{I}(\xc)+\frac1{2!}\mathcal{I}(\xc)}_{\bbit{3}{hs}+\bbit{3}{sh}}
   \underbrace{-\frac1{2!} \mathcal{K}(\xc)}_{\bbit{3}{hd}}
%\\&\qquad
%
 + \underbrace{3\times\frac1{3!} \mathcal{J}(\xc)}_{\bbit{3}{hss}+\cdots}
 + \underbrace{3\times\frac1{3!} \mathcal{K}(\xc)}_{\bbit{3}{hhs}+\cdots}
 \, .
\end{align*}
The auxiliary integrals $\mathcal{I}^{(1)}$, $\mathcal{J}$ and
$\mathcal{K}$ cancel as do the explicit $1/\epsilon$ poles in the
first line. The other contributions in \eqref{eq:nnnlocomb} are also
separately finite. Thus, after setting $d=4$ the explicit expressions
of the separately finite parts of \eqref{eq:nnnlocomb} are given by
\eqref{eq:nnnloparts} with
\begin{align}
\label{eq:nnnloidiv}
\bbit{3}{n+1}(\xc) &= \bbit{3}{f} \, ; &
\bbit{3}{n+2}(\xc) &= \bbit{3}{hf} \, ; &
\bbit{3}{n+3}(\xc) &= \bbit{3}{hhh} \, .
\end{align}
Comparing \eqref{eq:nnnloidiv} to \eqref{eq:nnloind1} and
\eqref{eq:nnloind2} reveals the pattern of how to extend beyond
N$^3$LO as done in Section~\ref{sec:nllo}.

%%%%%%%%%%%%%%%%%%%%%%%%%%%%%%%%%%%%%%%%%%%%%%%%%%%%%%%%%%%%%%%%%%%%%
\bibliographystyle{JHEP}
\bibliography{../muon_ref}{}

\end{document}